     \documentclass[twoside,journal]{IEEEtran}
    \usepackage{makecell}
    \usepackage{array}
    \usepackage{graphicx,amssymb,amsmath}
    \usepackage{multicol}
    \usepackage[noadjust]{cite}
    \usepackage{setspace}
    \usepackage{subfigure}
    \usepackage{graphicx}
    \usepackage{float}
    \usepackage {url}
    \usepackage{stfloats}
    \usepackage{amsthm,pifont}
    \usepackage{flushend}
    \usepackage{cases,subeqnarray}
    \usepackage{bm,multirow,bigstrut}
    \usepackage{amsmath, amsthm, amssymb}
    \usepackage{textcomp}
    \usepackage{latexsym,bm}
    \usepackage{booktabs}
    \usepackage{xcolor}
    \usepackage{mathtools}
    \usepackage{dsfont}
    \usepackage{extarrows}
    \usepackage{epsfig}
    \usepackage{epsfig}
    \usepackage{epstopdf}
    \usepackage[noend]{algpseudocode}
    \usepackage{algorithmicx,algorithm}

    \theoremstyle{plain}

    \theoremstyle{plain}

    \newtheorem{them}{Theorem}

    \usepackage{amsmath}
    
    \newtheorem{lemma}{Lemma}

    \IEEEoverridecommandlockouts
    
\begin{document}
    \title{Semantic-Aware Sensing Information Transmission for Metaverse: A Contest Theoretic Approach}

    \author{Jiacheng Wang, Hongyang Du, Zengshan~Tian, Dusit~Niyato,~\IEEEmembership{Fellow,~IEEE}, Jiawen~Kang, and Xuemin~(Sherman)~Shen,~\IEEEmembership{Fellow,~IEEE}
    \thanks {J.~Wang is with School of Communication and Information Engineering, Chongqing University of Posts and
    Telecommunications, Chongqing, China, and also School of Computer Science and Engineering, Nanyang Technological University, Singapore (e-mail: jcwang\_cq@foxmail.com).}

    \thanks{H.~Du and D. Niyato are with the School of Computer Science and Engineering, Nanyang Technological University, Singapore (e-mail: hongyang001@e.ntu.edu.sg, dniyato@ntu.edu.sg).}
    
    \thanks {Z.~Tian is with School of Communication and Information Engineering, Chongqing University of Posts and
    Telecommunications, Chongqing, China (e-mail:  tianzs@cqupt.edu.cn, corresponding author).}
    \thanks{J. Kang is with the School of Automation, Guangdong University of Technology, China (e-mail: kavinkang@gdut.edu.cn).}
    \thanks{X. Shen is with the Department of Electrical and Computer Engineering, University of Waterloo, Canada (e-mail: sshen@uwaterloo.ca).}

    }

    \maketitle
    \vspace{-1.5cm}
    \begin{abstract}
With the advancement of network and computer technologies, virtual cyberspace keeps evolving, and Metaverse is the main representative. As an irreplaceable technology that supports Metaverse, the sensing information transmission from the physical world to Metaverse is vital. Inspired by emerging semantic communication, in this paper, we propose a semantic transmission framework for transmitting sensing information from the physical world to Metaverse. Leveraging the in-depth understanding of sensing information, we define the semantic bases, through which the semantic encoding of sensing data is achieved for the first time. Consequently, the amount of sensing data that needs to be transmitted is dramatically reduced. Unlike conventional methods that undergo data degradation and require data recovery, our approach achieves the sensing goal without data recovery while maintaining performance. To further improve Metaverse service quality, we introduce contest theory to create an incentive mechanism that motivates users to upload data more frequently. Experimental results show that the average data amount after semantic encoding is reduced to about 27.87\% of that before encoding, while ensuring the sensing performance. Additionally, the proposed contest theoretic based incentive mechanism increases the sum of data uploading frequency by 27.47\% compared to the uniform award scheme.

    \end{abstract}
    \IEEEpeerreviewmaketitle
    \section{Introduction}
    In 1992, Neal Stephenson proposed the concept of Metaverse in its novel $Snow\ Crash$ for the first time. As an imagined virtual environment parallel to the physical world, in Metaverse, users can build their own avatars, analogous to their physical selves, to experience an alternate life~\cite{xu2022full}. Due to limited technology, resources, and other factors, at that time, Metaverse appeared to exist only in fantasy, which was out of reach. However, as time goes by, the economies have thrived and various technologies have evolved, bringing the once-imagined world into reality. In 2021, Meta (formerly Facebook) released Horizon Worlds~\cite{kraus2022facebook}, a free virtual reality online video game with an integrated game creation system, which allows users to build their own virtual characters and explore the space around them within the confines of their physical floor-space. In addition to giving people a better understanding of Metaverse, similar releases also mark the transformation of Metaverse's from fantasy to reality.
    
    As a novel type of Internet application and social form, Metaverse's transformation from a concept to reality is inextricably linked to innovative technologies. Digital twin-based data replication, for example, supports continuous data synchronization for Metaverse, that is, obtaining live data from the physical world and transmitting it to Metaverse to keep it updating in real time~\cite{el2018digital}.{\color{black}The 
   extended reality technologies can render Metaverse and create an immersive experience for users~\cite{xi2022challenges}. Moreover, blockchain technology, which is capable of tightly integrating the virtual world and the real world into the economic, social, and the identity system~\cite{ynag2022fusing}, is also essential.} 
    
    While eye-catching, the implementation of the above technologies rests on a fundamental premise, i.e., the data portraying the physical world can be effectively (accurately) and efficiently (fast) transmitted to Metaverse, which is of great challenge due to the large amount and variety of data. Taking the popular wireless sensing as an example, a pair of sensing devices can generate 3.072 megabytes of sensing data per second\footnote{Consider a transmitter equipped with one antenna sends signal with bandwidth of 80 Mhz and packet transmission rate of 500 Hz, and the receiver captures signal with three antennas, under the IEEE 802.11ac protocol. Then, the amount of sensing data produced per second can reach 3 × 256 × 500 × 2 × 4 bytes, where 256 is the number of Orthogonal Frequency Division Multiplexing (OFDM) subcarriers, 2 represents $I$ and $Q$ sampling.}. Even, this number could grow further as sensing granularity and the number of users increase, triggering not only an unbearable burden on data transmission and storage but also a non-negligible data processing delay, affecting the quality of service (QoS) of the MSP. For example, equipment providers may struggle to support high refresh rate and rendering due to storage and computing overhead. Therefore, the effective and efficient transmission of sensing data is a urgent issue that must be addressed.
    
    Semantic communication, which intends to extract the meaning of data, filter out the irrelevant information, and transmit it to the receiver, has natural advantages in reducing data storage and transmission load and improving data transmission efficiency~\cite{yang2022semantic}. Semantic communication is task-oriented, i.e., aims to convey the information relevant to the transmission goal. In image transmission, for instance, the information carried in the picture is essential, instead of the picture itself. With this in mind, semantic communication extracts information from the picture according to the transmission goal and gives it to the receiver to realize higher transmission efficiency. Therefore, if sensing data can be transmitted semantically, the burden of data communication and storage would be greatly reduced, laying the foundation for improving the QoS of MSP.
    
Inspired by this, in this paper, we propose a semantic-aware sensing data transmission framework, including a contest theory based incentive mechanism, to deliver the wireless sensing data, i.e., channel frequency response (CFR) power, from the physical world to Metaverse. As shown by the proposed framework in Fig.~\ref{img}, the transmitter focuses on data semantic coding and transmitting, the receiver works on data receiving, human activity recognition, and CFR power recovery to support other applications, while the contest theory based mechanism concentrates on encouraging the transmitter to upload sensing data at a higher frequency via award setting. Through such a collaboration, the amount of sensing data transmitted each time decreases significantly while the data upload frequency increases remarkably, providing support for MSP to further improve its QoS. The contributions of this paper are summarized as follows:
    \begin{figure*}[t]
    \centering
    \includegraphics[height=5.5cm]{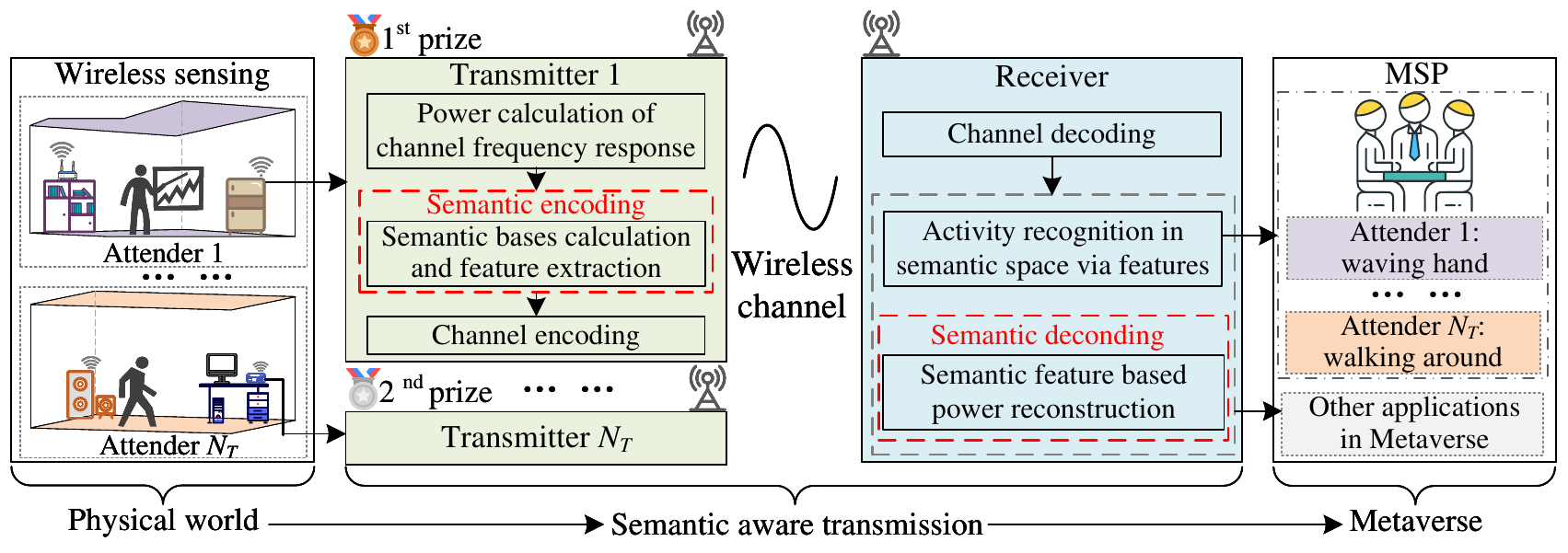} 
    \caption{The framework of the proposed system.} 
    \label{img} 
    \end{figure*}
   
    \begin{itemize}
    \item We define a novel semantic base to encode the information of human activity in the CFR. On this basis, the semantic features are further extracted and transmitted to receiver. Unlike transmitting the original CFR, our result shows the average data amount after semantic encoding is compressed to about $27.87\%$ of that before encoding. 
    
    \item We propose a semantic space, in which receiver realizes human activity recognition via the received semantic features, without the requirement for recovering the entire CFR power. We build an IEEE 802.11ac protocol based platform and collect data in real-world scenarios to verify the effectiveness of semantic coding and activity recognition in semantic space.
    
    \item We design a contest theory based strategy to incentivise the uploading of sensing data, to further improve the users experience in Metaverse. We derive the optimal award setting that maximises the total data uploading frequency, which directly affects the refresh rate of MSP, and investigate the advantages of semantic encoding in the contest. Numerical analysis proves and demonstrates the effectiveness of our proposed strategy.
    
    \end{itemize}
    \section{Related work}
    This section investigates the major works related to the system proposed in this paper.
  
    \subsection{Compressed Sensing}
    The channel state information (CSI) compression has always been a major topic in mobile wireless network. Previous attempts~\cite{xu2011feedback,xie2013adaptive,love2008overview} use quantization and general techniques to compress the CSI. For example, the authors in~\cite{xie2013adaptive} quantize CSI from time, frequency and numerical values domains, and adapt compression intensity to balance the capacity loss and overhead reduction. Other works~\cite{huang2009limited, carbonelli2007sparse} reduce the request rate and amount of CSI to achieve the compression. The authors in~\cite{huang2009limited} evaluate the required CSI repetition rate for multiple-input multiple-out put (MIMO) systems and provide a theoretical support for reducing CSI transmission in time domain. Besides, the sparse nature of the wireless channel is always used to underpin CSI compression~\cite{duarte2011structured, son2019analysis}. The authors in~\cite{son2019analysis} analyze CSI compression via the sparsity obtained by modeling feedback distortion and rate loss. CsiNet~\cite{wen2018deep} learns a transformation from CSI to a near-optimal number of representations via deep learning to realize CSI sensing and recovery. Moreover, CsiNet-LSTM~\cite{wang2018deep} improves the CSI recovery quality and trade-off between compression ratio and complexity by learning spatial structures from training samples.
    \subsection{Semantic Communications}
    The semantic communication was first introduced in \cite{ weaver1949mathematics}, then researchers present the semantic information theory \cite{carnap1952outline} and a universal semantic communication model \cite{bao2011towards}. Following that, the authors in \cite{juang2011quantification} develop a compression theory at the semantic level to reduce the amount of data. The authors in \cite{basu2014preserving} use the semantic redundancy and ambiguity to achieve compression. Using AI techniques, semantic communication is further developed both in technology and application. DeepSC \cite{xie2021deep} utilizes Transformer and transfer learning to maximize the capacity and minimize the errors in sentence meaning recovery, so as to perform better when the signal-to-noise ratio (SNR) is poor. Through the adversarial training and masked auto-encoder, the authors in \cite{hu2022robust} present the first end-to-end semantic communication system to improve the performance against noise. A Transformer based semantic communication system is presented in \cite{zhou2021semantic}, which introduces a circulation mechanism to transmit sentences more flexibly. Besides, semantic transmission of speech and images has also received extensive attention \cite{weng2021semantic, huang2021deep}.

    \subsection{Contest Theory}
    The contest is a game, in which contestants spend resources to win prizes with some probability. This definition can be applied to many situations to solve different problems. Early attempts, such as rent-seeking model~\cite{tullock1981rent}, all-pay auction~\cite{moulin1986game}, and others~\cite{broecker1990credit, anton1992coordination}, focused on economic problems. Besides economics, contest theory also plays a significant role in other areas. The authors in~\cite{lu2019multi} use contest to balance request and service among users, so as to solve the problem of low willingness of service in a service exchange application. {\color{black}In~\cite{vojnovic2015contest}, the authors use the contest theory in crowd sourcing to study the reward distribution strategies and estimate users' capabilities via the observed competition results.} The authors in~\cite{luo2015incentive} show requester can maximize contributions by rewarding only the top contributors, whereas the low-capability participants become risk-averse and are unwilling to join the contests. To handle this problem, the following work~\cite{luo2015crowdsourcing} introduces a lottery mechanism based on Tullock contest, so that every player has a positive chance of winning if they participate.
    
    Motivated by the above studies, we design a semantic transmission system for wireless sensing data. Our system uses the proposed semantic encoding to reduce the amount of sensing data that needs to be transmitted. Besides, we propose a competition based mechanism to stimulate the data upload frequency, which further improves the system overall performance.
    
    \section{System overview}
    \subsection{Wireless Sensing in Physical World}
    As shown in Fig. \ref{img}, without loss of generality, we use the activity recognition of users in virtual conference service as an example to illustrate our work. In the physical world, smart devices sense the attendee by transmitting and receiving signals and perform semantic encoding and other processing on the obtained sensing data. Here, the data refers to the CFR that is widely available in existing wireless communication systems and its power describes the physical channel fluctuation induced by attendee activity. Therefore, the CFR power, dented as {\small ${\left| {H\left( {f,t} \right)} \right|^2}$}, where $f$ and $t$ represent the frequency and time of CFR extraction, respectively, holds the potential for the pervasive sensing in the indoor environment \cite{wang2017device}. Once the CFR obtained, the transmitter needs to process it. However, there are two issues need to be addressed. 
    \begin{itemize}
    \item Due to the limited storage and computing resources, sensing devices need to send the {\small ${\left| {H\left( {f,t} \right)} \right|^2}$} via the transmitter module to the receiver, e.g., edge cloud nodes or fog nodes close to the sensing devices, for activity recognition. However, as mentioned previous, the sensing incurs a large amount of data, resulting in significant communication overheads.
    \item One receiver needs to serve multiple transmitters. Therefore, if all transmitter upload their data simultaneously to the receiver, the receiver will face significant costs due to the high demands on processing power and storage space.
    \end{itemize}
    These issues motivate us to develop a simple but effective semantic encoder to reduce the amount of data and the processing burden of receiver while maintaining activity recognition accuracy.
    \subsection{Semantic-aware Encoding at Transmitter}
    In this paper we focus on extracting and preserving semantic information related to attendee activity from CFR power, which allows us to transmit and store less data. Recall that the sinusoidal wave, the most natural representation of a signal, can be stored with only three parameters, i.e., amplitude, frequency, and initial phase. Although an infinite number of sinusoidal waves may be required to accurately describe a signal, we show later that a few waves are sufficient to maintain the CFR semantic properties required for human activity recognition. Therefore, we define semantic base as the sinusoidal wave, and semantic features as the corresponding parameters, including amplitude, frequency, and phase. Through the semantic bases, CFR can be semantically encoded. The theoretical support is presented in Section~\ref{eakfjha}. Given this, we consider the case that one receiver with $N_R$ antennas serves $N_T$ transmitters, as shown in Fig. \ref{img}. Each transmitter only needs to transmit semantic features to the receiver through the wireless channel, after CFR power calculation, semantic encoding, and channel encoding. By doing so, the transmission and storage resources are greatly reduced.
    \subsection{Activity Recognition at Receiver with Contest Incentive}
    With respect to the transmitter, the receiver first conducts channel decoding to extract semantic features from the received signal. The features are directly related to the activities of the conference attendees. Therefore, the $k$-nearest neighbor (kNN) is applied to classify the semantic features in semantic space (detailed in Section IV-C) to achieve activity recognition. After that, the recognition results are fed to the MSP to complete the virtual conference function in Metaverse. Besides, receiver can restore the entire CFR power by combining sine waves described by the semantic features, so as to support other services in Metaverse.
    
    The receiver serves multiple transmitters at the same time, each of which has different sensing, processing, and data transmission capabilities. As mentioned previously, the data transmission rate directly influences the refresh rate and rendering of the MSP, which further affects the user experience. Therefore, we introduce a reward mechanism (in Section IV-D) for transmitters of the service framework according to the contest theory. To win the reward, each transmitter needs to upload sensing data with a higher frequency. Therefore, with this incentive, the overall rendering and refresh effects at the MSP are boosted, which further enhances the user experience.
    \section{System Model}
    \subsection{Wireless Sensing and Semantic-Aware Encoding}\label{eakfjha}
    In this part, the theoretical support of semantic-aware encoding is introduced. As the framework shown in Fig.~\ref{img}, various wireless IoT devices in the physical world are used to sense the conference attendee by sending and receiving wireless signals. Since the surrounding objects, the transmitted signal reaches at the receiver via multiple propagation paths. {\color{black}Assuming there are $L$ different propagation paths,} then the CFR of the wireless channel can be denoted as 
    \begin{align}\label{eq1}
    &H\left( {f,t} \right) = \sum\limits_{l = 1}^L {{a_l}\left( {f,t} \right)\exp \left( { - j2\pi f\frac{{{d_l}\left( t \right)}}{c}} \right)}\notag\\
    &\times\exp \left( { - j2\pi \left( {f\left( {{\varepsilon _1} + {\varepsilon _2}} \right) + {\varepsilon _3}} \right)} \right),
    \end{align}
    where ${a_l}\left( {f,t} \right)$ represents the attenuation and the initial phase of the $l$-th propagation path, $\exp \left( { - j2\pi f{{{d_l}\left( t \right)} \mathord{\left/ {\vphantom {{{d_l}\left( t \right)} c}} \right. \kern-\nulldelimiterspace} c}} \right)$ is the phase introduced by propagation delay corresponding to the distance ${d_l}\left( t \right)$, $c$ is the signal propagation speed in the air, and ${\varepsilon _1}$, ${\varepsilon _2}$, and ${\varepsilon _3}$ are the time offset introduced by the symbol timing offset (STO), sampling frequency offset (SFO), and carrier frequency offset (CFO), respectively. {\color{black}As shown in \eqref{eq1}, the change of ${d_l}\left( t \right)$ triggers a phase shift to the signal corresponding to the $l$-th propagation path.} In the indoor environment, the propagation path length of reflections induced by static objects, such as walls and furniture, are constant, while the path length of the conference attendee induced reflection varies with time. Consequently, the propagation paths can be categorized into static and dynamic parts. Leveraging this fact, we can rewrite \eqref{eq1} as
    \begin{align}\label{fml2}
    H\left( {f,t} \right) = \exp \left( { - j2\pi \varepsilon } \right) \times \left( {{H_s}\left( {f,t} \right) + {H_d}\left( {f,t} \right)} \right),
    \end{align}
    where $\varepsilon  = f\left( {{\varepsilon _1} + {\varepsilon _2}} \right) + {\varepsilon _3}$, constant vector ${H_s}\left( {f,t} \right)$ is the sum of static reflections, and 
    \begin{align}\label{fml3}
    {H_d}\left( {f,t} \right) = \sum\limits_{{l_d} \in {P_d}} {{a_{{l_d}}}\left( {f,t} \right)} \exp \left( { - j2\pi f{{{d_{{l_d}}}\left( t \right)} \mathord{\left/
    {\vphantom {{{d_{{l_d}}}\left( t \right)} c}} \right.
    \kern-\nulldelimiterspace} c}} \right)
    \end{align}
    is the sum of dynamic reflections, ${P_d}$ is the set of the dynamic reflections. Without loss of generality, we can assume that the ${l_d}$-th propagation path is introduced by the conference attendee and path length changes at a constant rate of ${v_k}$ in a short period of time. Then, we have
    \begin{align}\label{fml4}
    {H_d}\left( {f,t} \right) = \sum\limits_{{l_d} \in {P_d}} {{a_{{l_d}}}\left( {f,t} \right)} \exp \left( {{{ - j2\pi f\left( {{d_{{l_d}}}\left( 0 \right) + {v_k}t} \right)} \mathord{\left/
    {\vphantom {{ - j2\pi f\left( {{d_{{l_d}}}\left( 0 \right) + {v_k}t} \right)} c}} \right.
    \kern-\nulldelimiterspace} c}} \right),
    \end{align}
    where ${d_{{l_d}}}\left( 0 \right)$ is the initial path length of attendee induced reflection. Apparently, the phase of the CFR contains information about ${v_k}$, which has been proven to be directly related to the activity of the human body \cite{wang2017device}. Due to the phase error $\varepsilon  = f\left( {{\varepsilon _1} + {\varepsilon _2}} \right) + {\varepsilon _3}$, it is hard to directly calculate ${v_k}$ from the CFR. Fortunately, besides the phase, the ${v_k}$ manifests itself via the power change of CFR, which can be illustrated via the example in Fig. 2. One can see from the figure that the overall CFR power undergoes a significant decrease after ${H_d}$ shift to ${H'_d}$, {\color{black}as evidenced by the shortening of CFR' in comparison with CFR, even if the power remains unchanged.}
    \begin{figure}[t]
    \centering
    \includegraphics[height=4cm]{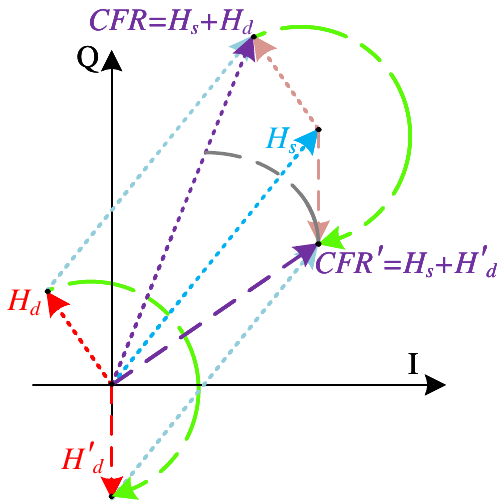} 
    \caption{The impact of the ${H_d}$ phase on the overall CFR power.} 
    \end{figure}
    Therefore, we calculate the CFR power and determine the semantic base by analyzing the relationship between power and ${v_k}$. Concretely, according to the  \eqref{fml2},  \eqref{fml3}, and  \eqref{fml4}, we have
    \begin{equation}\label{fml5}
    {\left| {H\left( {f,t} \right)} \right|^2} = {H_D} + {H_S} + {H_{DC}},
    \end{equation}
    where
    \begin{align}
    &{H_D} = \sum\limits_{{l_d} \in {P_d}} {2\left| {{H_s}\left( {f,t} \right){a_{{l_d}}}\left( {f,t} \right)} \right|} \notag\\
    &\times
    \cos \left( {\frac{f}{c}\left( {2\pi {v_k}t + 2\pi {d_{{l_d}}}\left( 0 \right)} \right) + {\varphi _{s{l_d}}}} \right)
    \end{align}
    is the cross-term,
    \begin{align}
   & {H_S} = \sum\limits_{{l_d},k \in {P_d},{l_d} \ne k} {2\left| {{a_{{l_d}}}\left( {f,t} \right){a_k}\left( {f,t} \right)} \right|} 
    \notag\\
    &\times\!
    \cos \!\left(\! {\frac{f}{c}2\pi \left( {\left( {{v_k}\! -\! {v_l}} \right)t \!+\! {d_{{l_d}}}\!\left( 0 \right) \!- \!{d_k}\!\left( 0 \right)} \right) + {\varphi _{{l_d}k}}} \!\right)\!,
    \end{align}
    is the self-term,
    \begin{equation}
    {H_{DC}} = \sum\limits_{{l_d} \in {P_d}} {{{\left| {{a_{{l_d}}}\left( {f,t} \right)} \right|}^2} + {{\left| {{H_s}\left( {f,t} \right)} \right|}^2}} ,
    \end{equation}
    is the DC-term, and ${\varphi _{s{l_d}}}$ and ${\varphi _{{l_d}k}}$ are the initial phase. From the above derivation, the following conclusions can be drawn. 
    \begin{itemize}
    \item The power of CFR consists of three parts. Sorted by the power in a descending order, these components are DC term, cross-term, and self-term. Excepting the DC term, the rest two terms are composed of multiple sinusoidal waves.
    \item The frequencies of the sinusoidal waves that make up cross-term and self-term are determined by the ${v_k}$, which, as mentioned previously, is directly related to the activity of human body. Moreover, the overall initial phases of these two terms are constants, which are determined by ${d_{{l_d}}}\left( 0 \right)$, ${\varphi _{s{l_d}}}$, ${d_k}\left( 0 \right)$, and ${\varphi _{{l_d}k}}$.
    \end{itemize}
    
    From the foregoing derivation and conclusions, it can be seen that the CFR power of different activities can be represented by the sum of a constant component and a set of sinusoidal waves. 
    
    {\color{black}According to these observation, we first employ principal component analysis (PCA) to denoise} {\color{black}the obtained CFR power \footnote{{\color{black}There are two main reasons for choosing PCA. First, the CSI contains burst noise introduced by internal state transitions. For cancelling such noise, PCA outperforms the traditional filtering-based algorithms \cite{wang2017device}. Second, through matrix decomposition and mapping, PCA can convert the CFR matrix (corresponding to the subcarrier and time domain) into vectors, reducing the data dimension, while cancelling the noise and preserving the signal characteristics.}}} {\color{black}and eliminate the constant component by subtracting the mean of the power\footnote{{\color{black}The DC term is a constant that does not contain any activity-related information, so it can be removed by subtracting the mean of the CFR power.}}. After that, a series of sinusoidal waves with different amplitudes, frequencies, and initial phases are selected to fit the CFR power to achieve semantic encoding. Here, the selected waves are defined as the semantic bases of the CFR and the corresponding wave parameters are the semantic features.} This semantic encoding process can be achieved via the Levenberg-Marquardt (LM) algorithm \cite{gao2021channel} and we summarize it in Algorithm 1, where $\#(Iteration)$ denotes the number of iterations, $order$ means the number of bases required for semantic encoding, $A$, $F$, and $\theta$ represent the amplitude, frequency, and initial phase of the semantic bases, respectively, $T_1$ is the fitting error threshold, $T_2$ is the threshold for the number of iterations, $\left| {H\left( {f,t} \right)} \right|_i^2$ is the $i$-th element of CFR power, $f{\left( {A,{\rm{ }}F,{\rm{ }}\theta } \right)_i}$ is the $i$-th element of fitted value, and $I$ is length of CFR for semantic encoding.
    
    \begin{algorithm}[t]
    \caption{Semantic-aware encoding.} 
    \label{Fills}
    \hspace*{0.02in} {\bf Input:}
    The original CFR obtained by sensing the physical world \\
    \hspace*{0.02in} {\bf Output:} The amplitude, frequency, and initial phase of semantic basis.
    \begin{algorithmic}[1]
    \State    $\#(Iteration)$=0; $order$=0;
    \State The CFR power calculation $\left| {H\left( {f,t} \right)} \right|_i^2,i = 1, \ldots , I$  
    \State The PCA based denoising and ${H_{DC}}$ removal
    \State Fast Fourier transform based order estimation and updating
    \State Initializing $A$, $F$, and  $\theta$
    \State Levenberg-Marquardt (LM) based fitting: $\mathop {\min }\limits_{A,{\rm{ }}F,{\rm{ }}\theta } {\rm{ }}{\sum\limits_{i = 1}^I {\left[ {\left| {H\left( {f,t} \right)} \right|_i^2 - f\left( {A,{\rm{ }}F,{\rm{ }}\theta } \right)} \right]} ^2}$
    \State Fitting error calculation
    \While{Fitting error $>$ $T_1$}
    \State $order$ =$order$ +1 and $\#(Iteration)$= $\#(Iteration)$+1
    \State Updating $A$, $F$, and  $\theta$
    \State LM based fitting
    \State Fitting error updating 
    \If {$\#(Iteration)$ $>$ $T_2$}
    \State break
    \EndIf			
    \EndWhile
    \State \Return {$\left[ {{A_r},{\rm{ }}{F_r},{\rm{ }}{\theta _r}} \right],{\rm{ }}r = 1:order$}.
    \end{algorithmic}
    \end{algorithm}
    
\begin{figure*}[t!]
    \centering
    \includegraphics[height=12cm]{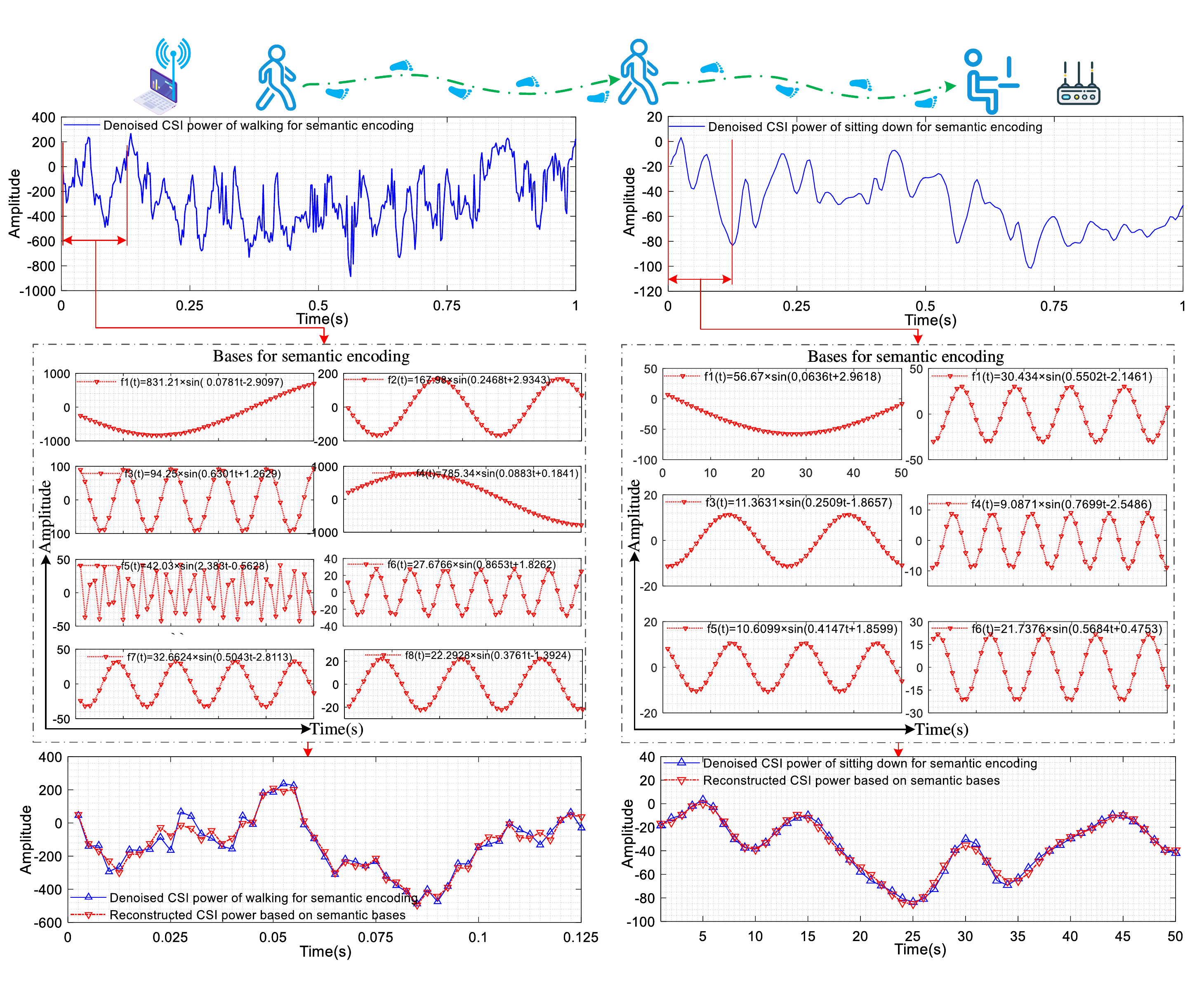} 
    \caption{An example of semantic encoding for walking and sitting.} 
    \label{exp ecd}
    \end{figure*}

    Taking the power of CSI (a sampled version of CFR collected via the Nexmon toolkit~\cite{gringoli2019free}) for two common activities, including walking and sitting, as an example, Fig.~\ref{exp ecd} presents the semantic encoding results. From the original CSI power and encoding results, one can see that walking contains richer components than sitting. Thus, eight semantic bases are used to encode it and the corresponding the semantic features, include the amplitude, frequency, as well as the initial phase, are shown as f1 to f8 in the figures. Relatively speaking, sitting down holds fewer components than walking. Therefore, six semantic bases are sufficient to realize semantic encoding, and the encoding outcome looks more impressive than walking. 
    
    So far, the rationality of the semantic bases and the feasibility of semantic encoding are proved by the theoretical derivation and the encoding examples. {\color{black}Clearly, semantic encoding reduces the amount of data that needs to be transmitted each time, while preserving the activity related information, providing more flexibility in increasing data transmission frequency and activity recognition.} To translate the gains obtained in semantic encoding into QoS improvement, we define the semantic space, in which the activity recognition is realized without reconstructing the CFR power. Moreover, an incentive mechanism is proposed in Section IV-C to increase the transmission frequency, providing support for boosting the refresh rate for MSP.  
    \subsection{Wireless Network}\label{fading}
    Note that the wireless environment between the transmitter and receiver could impact the performance of the system. For example, the bit error probability (BEP) caused by wireless transmission can affect the accuracy of the activity recognition. To investigate this, we consider a generalised network architecture, where the distance between the transmitter and the receiver is $D_w$, and the large-scale path loss exponent is $\alpha_w$. Considering the small-scale shadow fading and multi-path fading, a generalized small-scale fading model is used to characterize the fluctuations in the amplitude of the transmitted signal. {\color{black}Since the wireless environment is dynamic and volatile, we use $H$-fading channel model~\cite{jeong2015h}, which includes most of typical models such as Rayleigh, Nakagami-$m$, Weibull, $\alpha$-$\mu$, $N$*Nakagami-$m$, generalized $K$-fading, and Weibull/gamma fading as its special cases. Specifically, the $H$-fading can be converted to the various commonly used small-scale fading models as \cite[Table 1]{kong2019physical}.} The PDF of $H$-fading is given as \cite{kong2019physical}
    \begin{equation}
    {f_X}\left( x \right) = \kappa H_{p,q}^{m,n}\left[ {\lambda x\left| {\begin{array}{*{20}{l}}
    {{{\left( {{a_i},{A_i}} \right)}_{i = 1:p}}} \\ 
    {{{\left( {{b_l},{B_l}} \right)}_{l = 1:q}}}
    \end{array}} \right.} \right]
    \end{equation}
    where $H \, \substack{ m , n \\ p , q}(\cdot)$ is the Fox's  $H$-function \cite[eq. (1.2)]{mathai2009h}, $ \lambda  > 0 $ and $\kappa$ are the constants and satisfy $ \int_0^\infty  {{f_X}} \left(x\right)d{x} = 1 $, ${{{\left( {{a_i},{A_i}} \right)}_{i = 1:p}}}$ means $ {\left( {{a_i},{A_i}} \right)_{i = 1:p}}{=}\left( {{a_1},{A_1}} \right), \ldots ,\left( {{a_p},{A_p}} \right) $. Assuming that maximum-ratio combining reception in the $n^{\rm th}$ transmitter~\cite{telatar1999capacity}, the output SNR at the receiver, i.e., $Z_n$, can be expressed as the sum of the individual branches~\cite{rahama2018sum}. Thus, we have $Z_n = X_1+\dots+X_{N_R}$, where $X_{j}$ $\left(j=1,\ldots,N_R\right)$ denotes the SNR between the transmitter and the $j^{\rm th}$ antenna in the receiver. The probability density function (PDF) and cumulative distribution function (CDF) of $Z_n$ are derived as in~\cite[eq. (8)]{rahama2018sum} and \cite[eq. (9)]{rahama2018sum}, respectively.
    
    \subsubsection{Data Rate Evaluation}
    The ergodic data rate (or Shannon capacity) for the $n^{\rm th}$ transmitter, which is known to be the maximum data rate that the channel can support, is defined as
    \begin{equation}\label{fdeklahjf}
    C_{\rm n} = B_n \int_0^\infty  {{{\log }_2}} (1 + \gamma ){f_{Z_n} }(\gamma )d\gamma, 
    \end{equation}
    where {$B_n$} is the available channel bandwidth of the $n^{\rm th}$ transmitter. By substituting \cite[eq. (8)]{rahama2018sum} into \eqref{fdeklahjf}, the data rate can be derived as \cite[eq. (28)]{rahama2018sum}.

    \subsubsection{Bit Error Rate (BER) Evaluation}
    The average BEP for the $n^{\rm th}$ transmitter under a variety of modulation formats is given by~\cite{tse2005fundamentals}
    {\small \begin{equation}\label{BEREQUATION}
    E_{\rm n} = \int_0^\infty  {\frac{{\Gamma \! \left( {{\tau _2},{\tau _1}\gamma } \right)}}{{2\Gamma \! \left( {{\tau _2}} \right)}}{f_{Z_n}}\left( \gamma  \right)d\gamma },
    \end{equation}}\noindent
    where ${\tau _1} $ and ${\tau _2}$ are modulation-specific parameters for several modulation and detection scheme combinations, respectively, $ {{{\Gamma \! \left( {{\tau _2},{\tau _1}\gamma } \right)}}/{{2\Gamma \! \left( {{\tau _2}} \right)}}} $ represents the conditional bit-error probability, different modulation and detection scheme combinations with corresponding parameters ${\tau_1}$ and ${\tau_2}$ are shown in Table \ref{parameters}, and $\Gamma \! \left( { \cdot , \cdot } \right)$ denotes the upper incomplete Gamma function~\cite[eq. (8.350.2)]{gradshteyn2007}.
    Substituting \cite[eq. (8)]{rahama2018sum} into \eqref{BEREQUATION}, we can obtain the close-form of BER as \cite[eq. (25)]{rahama2018sum}. With the modeling ability of the $H$ fading model for various channel environments, we analyze the impact of bit errors on semantic information transmission in Section~\ref{fealjk}.
    \begin{table}
    \caption{Different modulation and detection scheme combinations with corresponding parameters ${\tau_1}$ and ${\tau_2}$.}
    \label{parameters}
    \centering
    \renewcommand{\arraystretch}{1}
    {\small\begin{tabular}{|m{0.5cm}<{\centering}|m{0.5cm}<{\centering}|m{6.5cm}<{\centering}|}
    \toprule
    \hline
    ${\tau_1}$ & ${\tau_2}$ & \textbf{Modulation \& Detection Scheme} \\
    \hline
    $0.5$ & $0.5$ & Orthogonal coherent binary frequency-shift keying (BFSK) scheme \\
    \hline
    $1$ & $0.5$ & Antipodal coherent binary phase-shift keying (BPSK) scheme \\
    \hline
    $0.5$ & $1$ & Orthogonal non-coherent BFSK scheme \\
    \hline
    $1$ & $1$ & Antipodal differentially coherent BPSK (DPSK) scheme \\
    \hline
    \bottomrule
    \end{tabular}}
    \end{table}
    
    Assuming that the number of bits generated by wireless sensing for each time is $D$, then the time required for the $n^{\rm th}$ transmitter to send the sensing data to receiver is $D/C_n$, and the average number of error bits is $E_{\rm n} D$. The amount of raw data generated by wireless sensing is relatively large as discussed above. Fortunately, the proposed semantic encoding greatly reduces the amount of data, enabling it to sense and upload data to complete recognition multiple times per second, which improves synchronization between the physical world and Metaverse.
    \subsection{Semantic-Aware Activity Recognition}
\begin{figure}[t]
    \centering
	    \includegraphics[height=5cm]{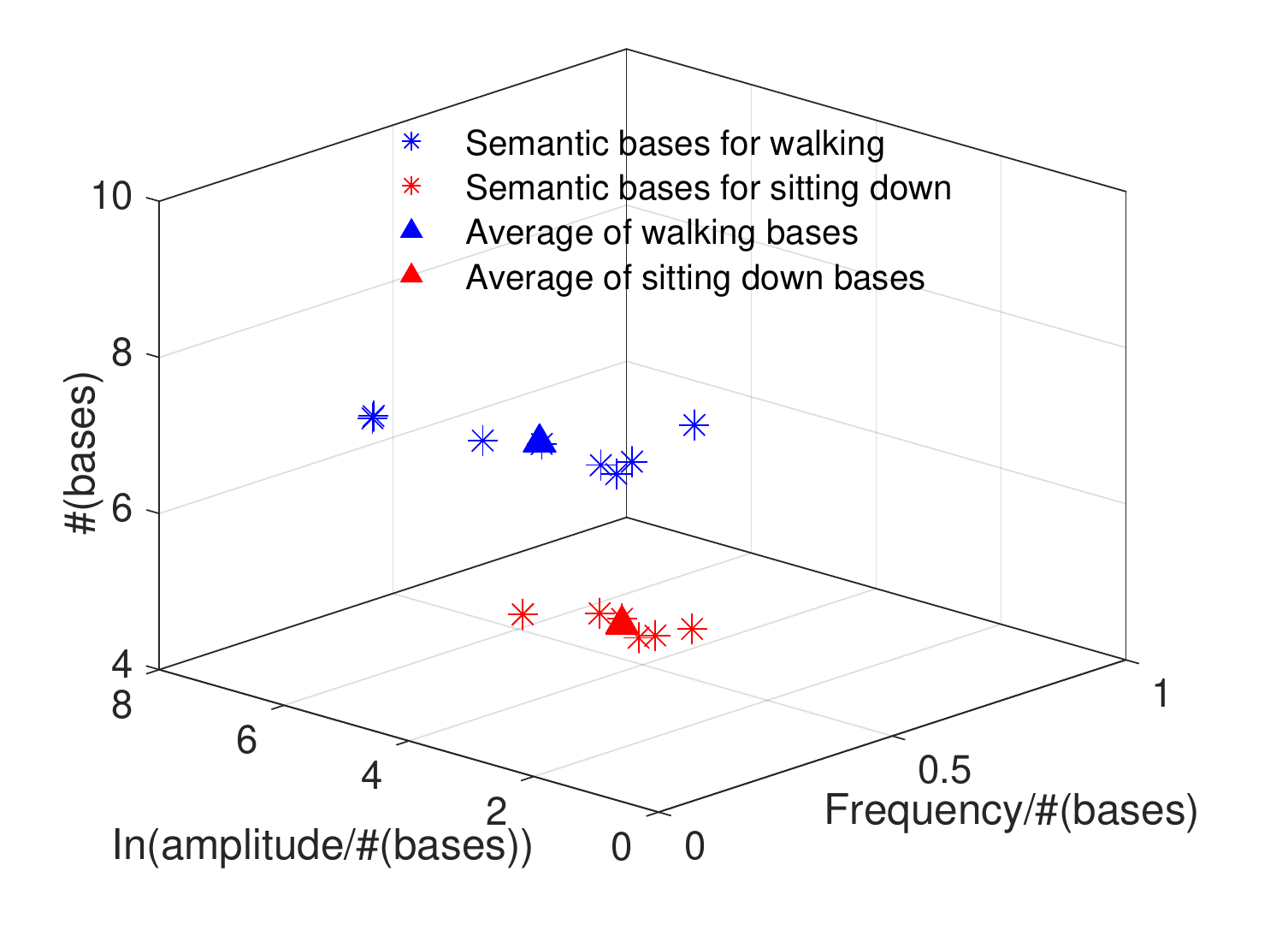} 
    \caption{The distribution of semantic features in defined 3D semantic space.} 
    \label{MP exp} 
    \end{figure}

\begin{figure*}[htbp]
    \centering
    \subfigure[]{
    \begin{minipage}[t]{0.26\linewidth}
    \centering
    \includegraphics[width=4cm]{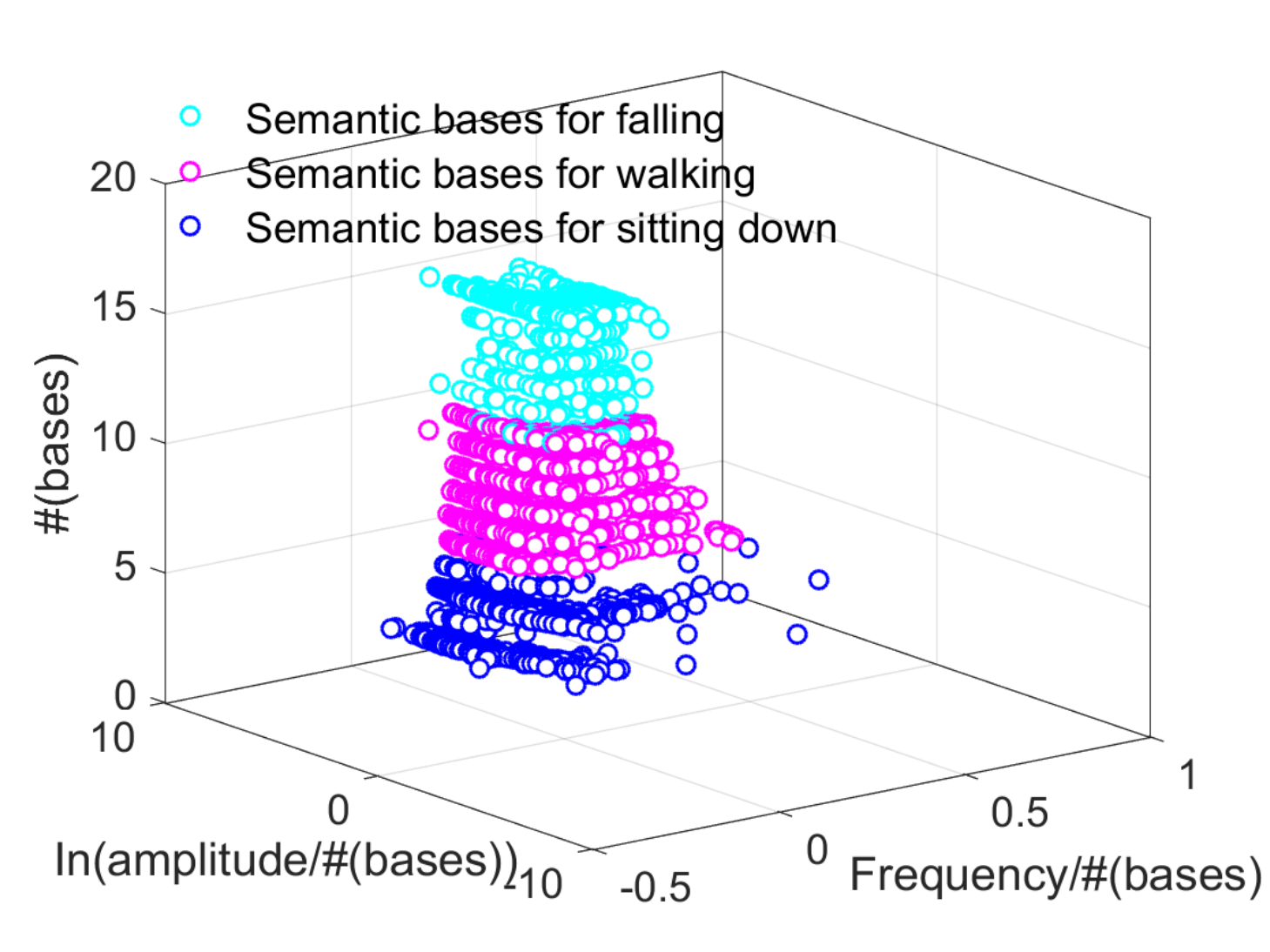}
    \end{minipage}%
    }%
    \subfigure[ ]{
    \begin{minipage}[t]{0.22\linewidth}
    \centering
    \includegraphics[width=4.5cm]{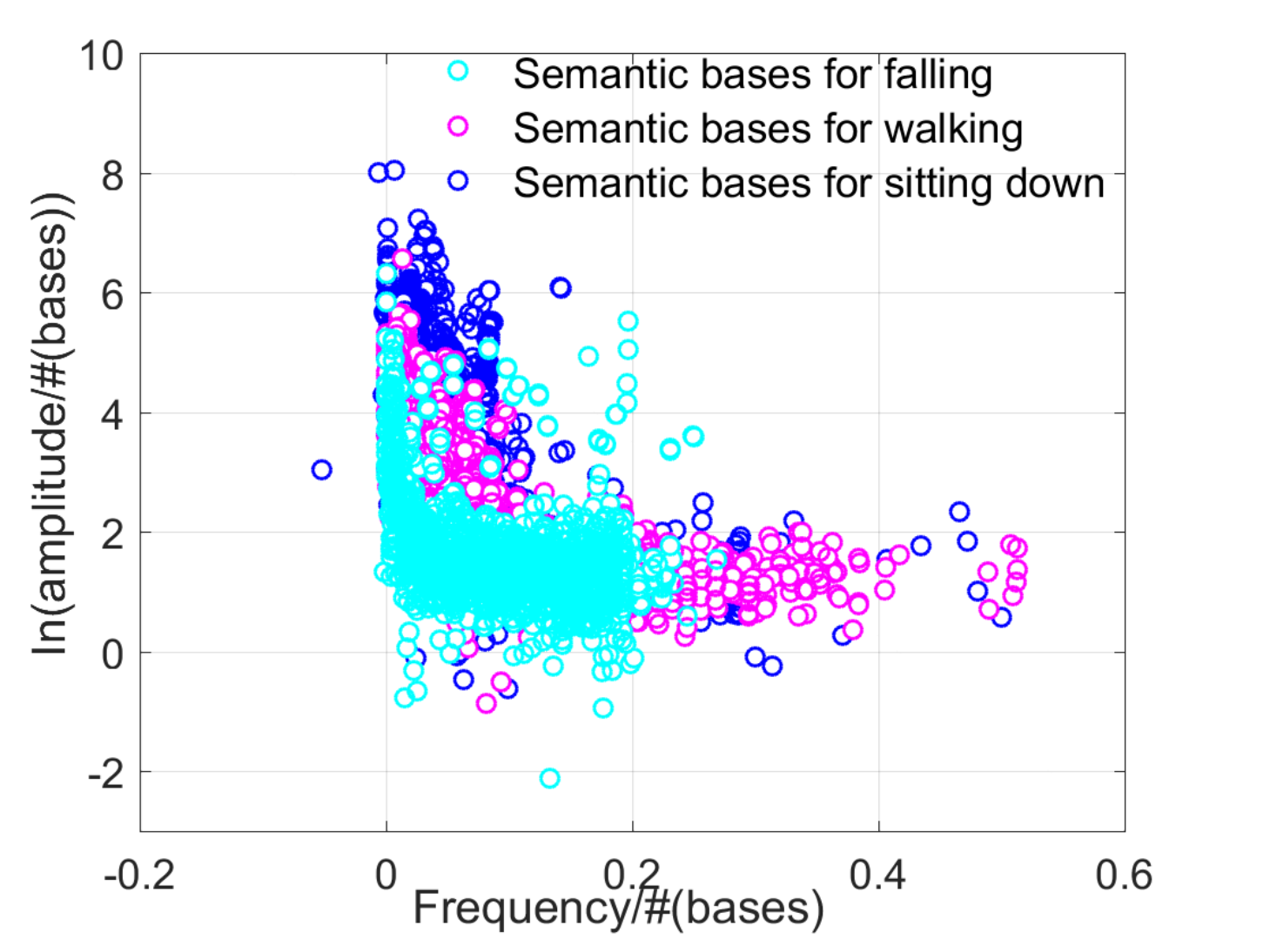}
    \end{minipage}%
    }%
    \subfigure[ ]{
    \begin{minipage}[t]{0.22\linewidth}
    \centering
    \includegraphics[width=4.5cm]{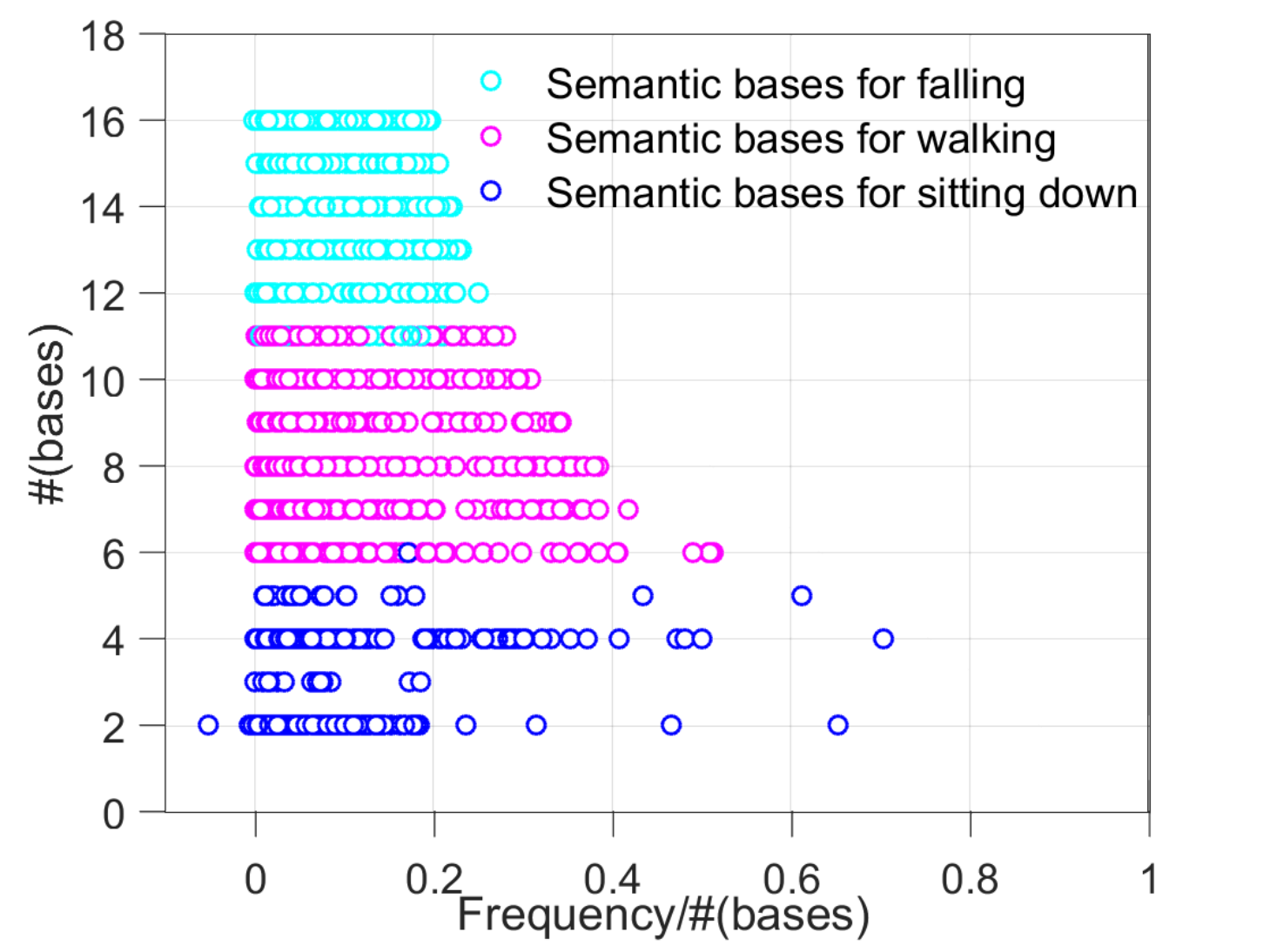}
    \end{minipage}
    }%
    \subfigure[]{
    \begin{minipage}[t]{0.26\linewidth}
    \centering
    \includegraphics[width=4.5cm]{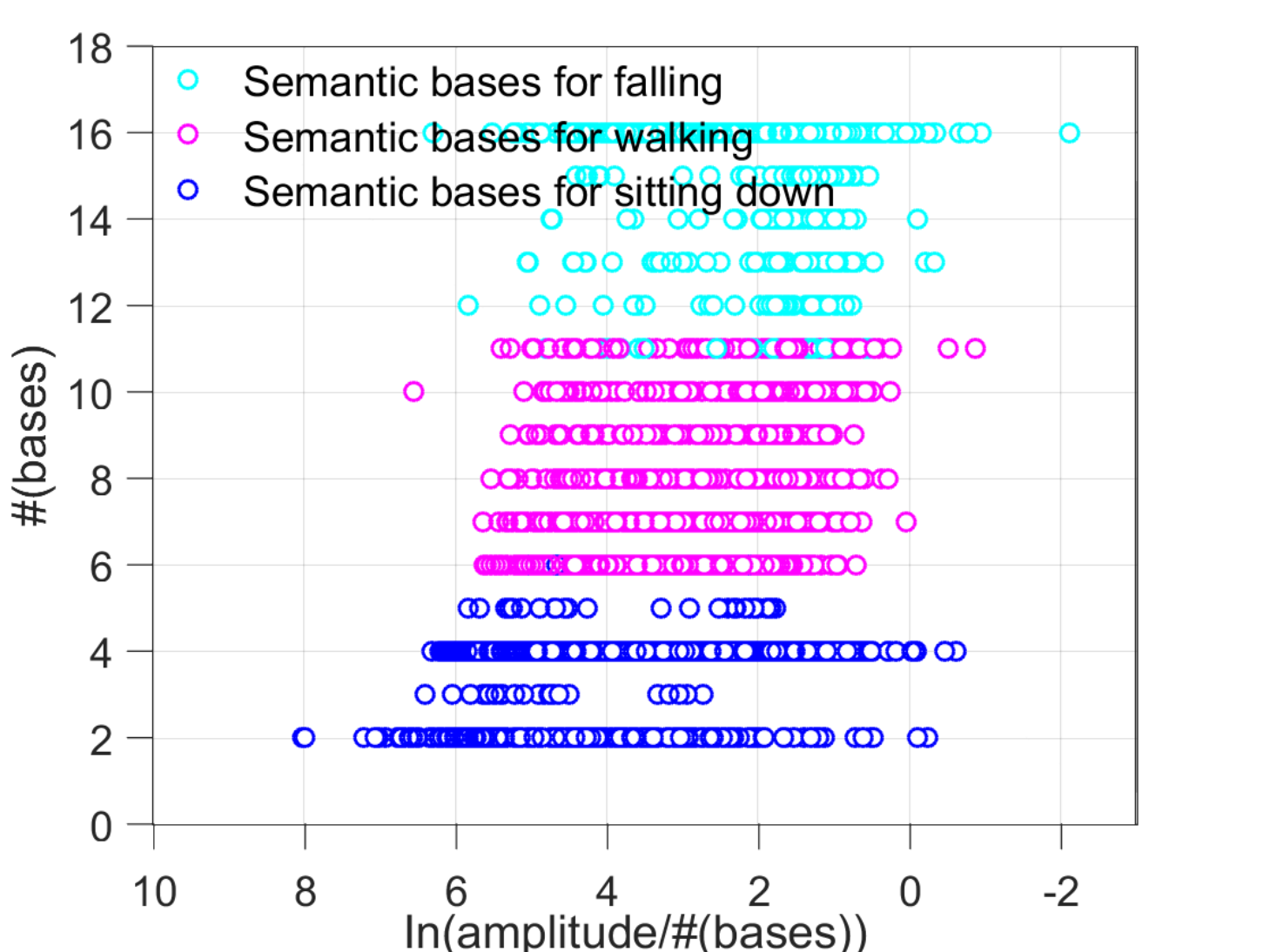}
    \end{minipage}
    }%
    \centering
    \caption{The distribution of features in the 3D semantic space. }
    \label{PITS} 
    \end{figure*}

\begin{figure*}[htbp]
    \centering
    \subfigure[]{
    \begin{minipage}[t]{0.26\linewidth}
    \centering
    \includegraphics[width=5cm]{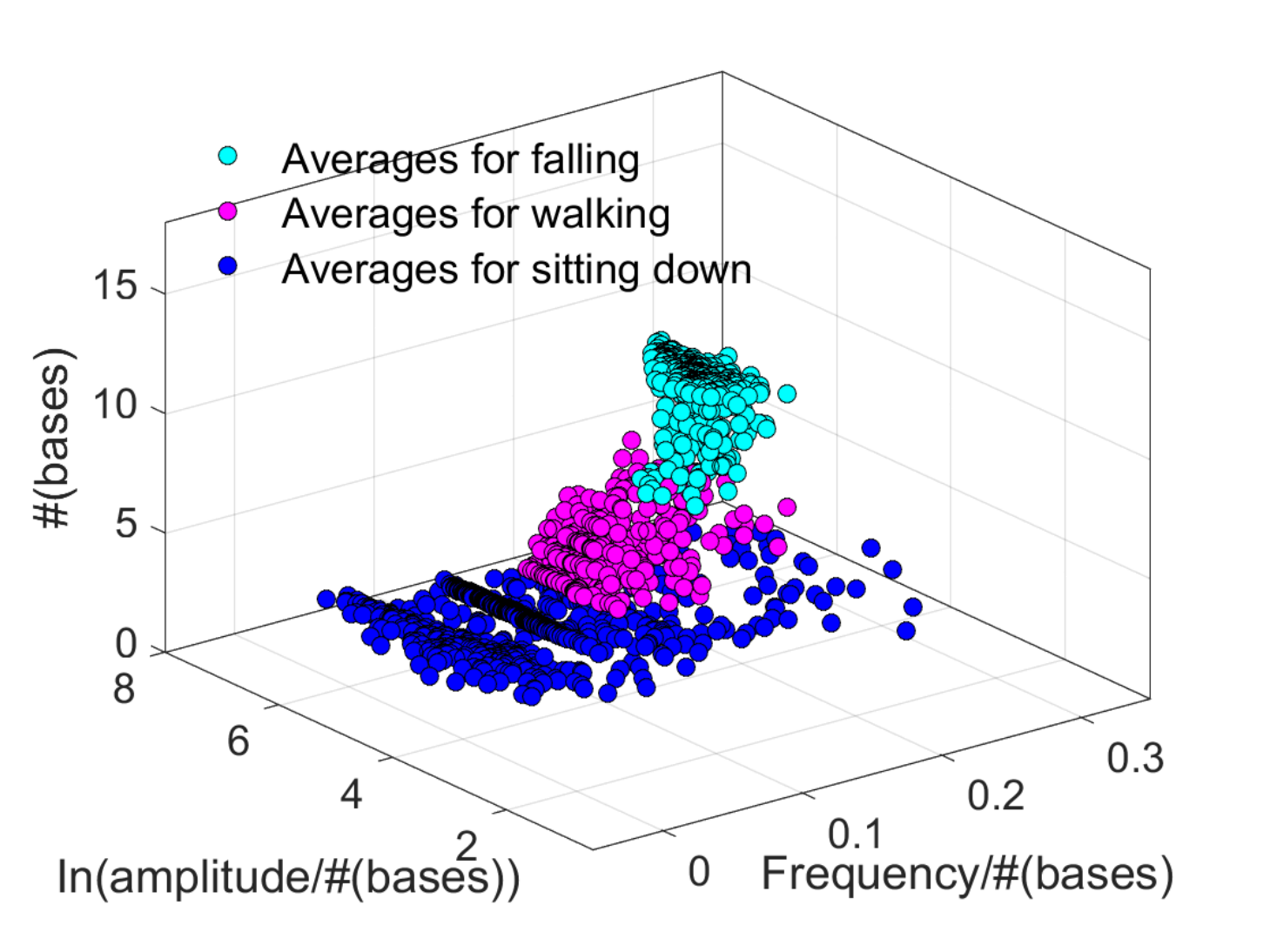}
    \end{minipage}%
    }%
    \subfigure[ ]{
    \begin{minipage}[t]{0.22\linewidth}
    \centering
    \includegraphics[width=4.5cm]{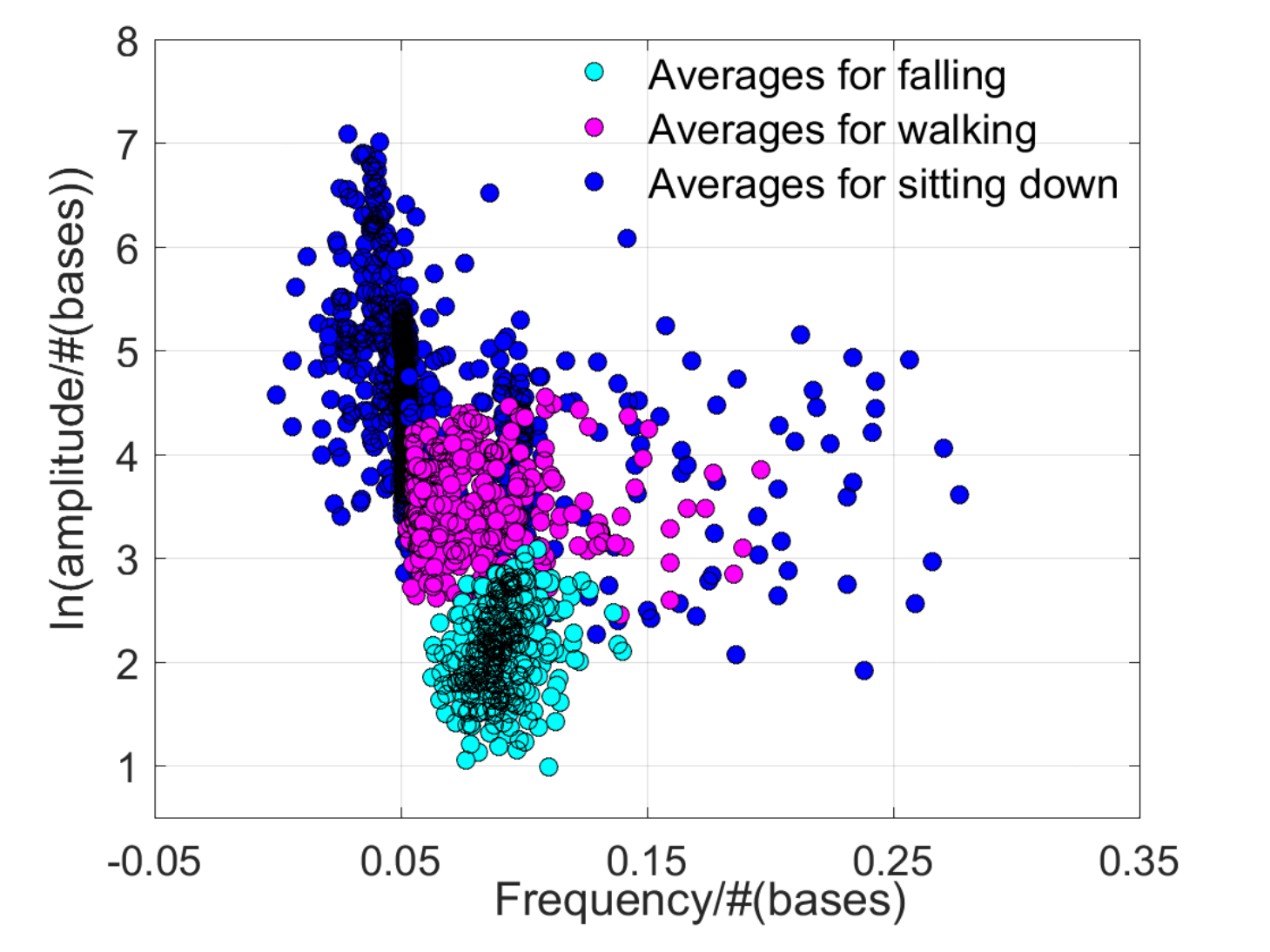}
    \end{minipage}%
    }%
    \subfigure[ ]{
    \begin{minipage}[t]{0.22\linewidth}
    \centering
    \includegraphics[width=4.5cm]{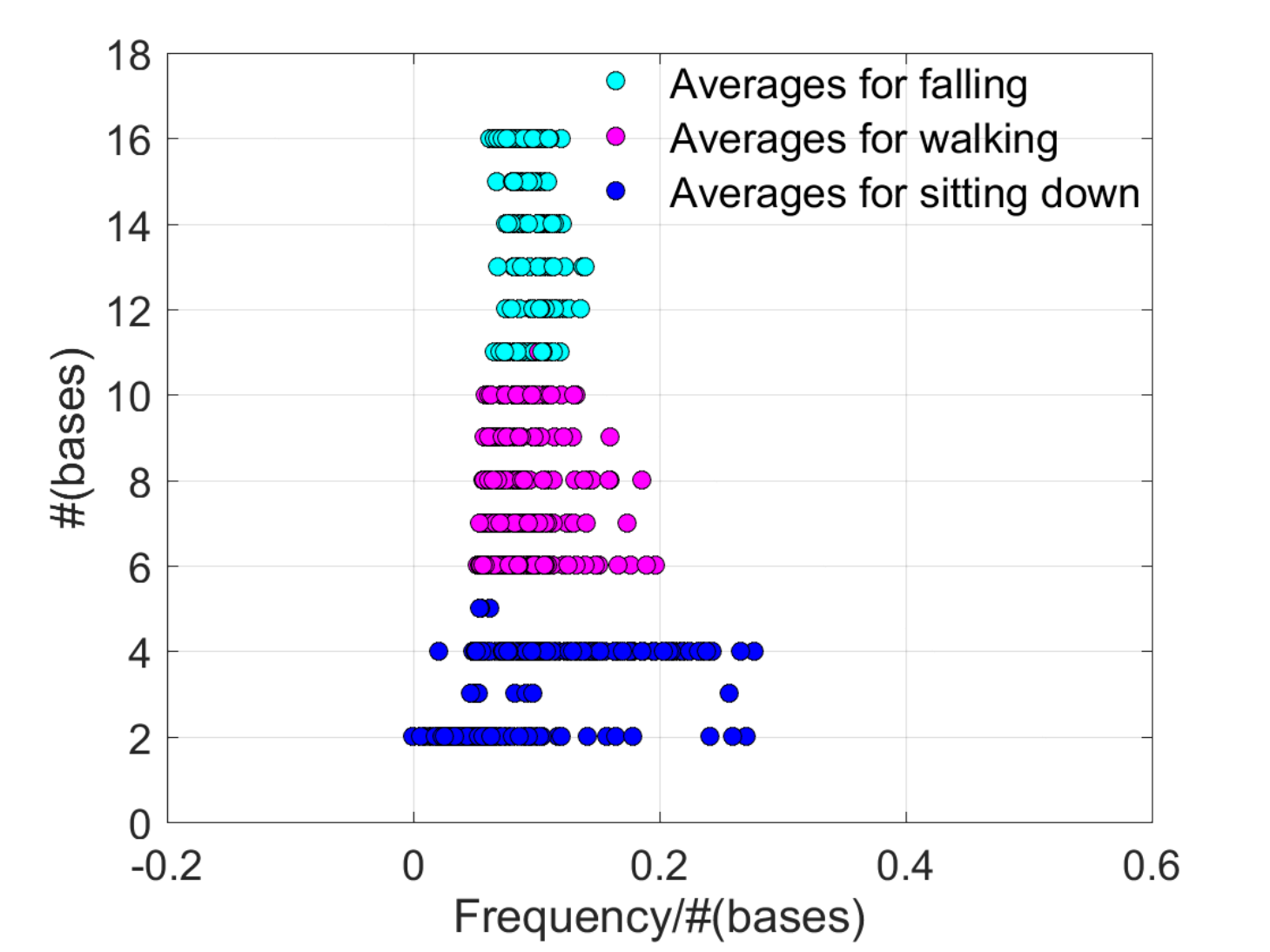}
    \end{minipage}
    }%
    \subfigure[]{
    \begin{minipage}[t]{0.26\linewidth}
    \centering
    \includegraphics[width=4.5cm]{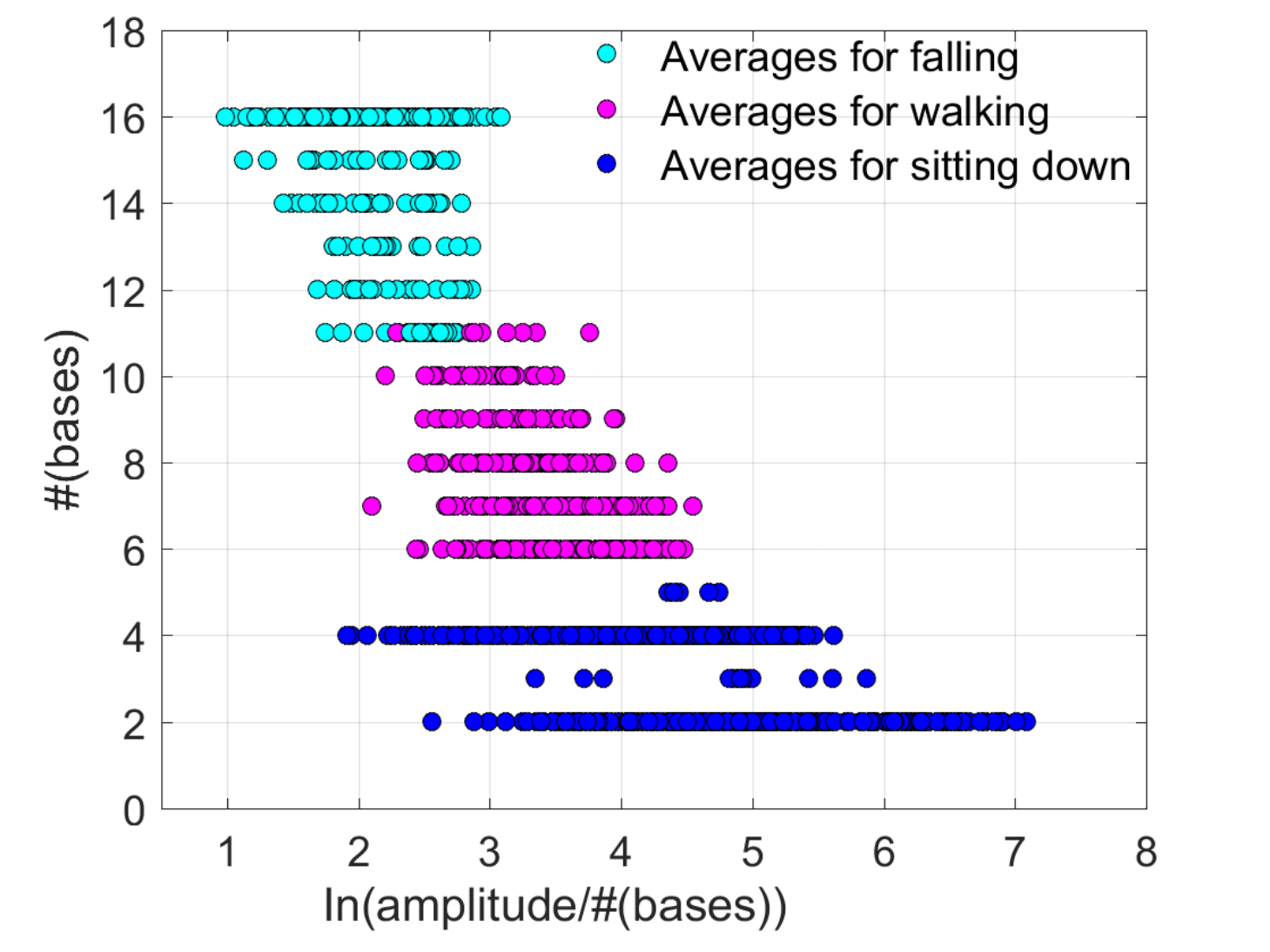}
    \end{minipage}
    }%
    \centering
    \caption{The distribution of features averages in the 3D semantic space. }
    \label{PITSAVRG} 
    \end{figure*}

    After receiving the signal, the receiver conducts channel decoding to obtain the semantic features, including the amplitude, frequency, and initial phase of the sine waves, through which the original CFR power can be recovered. Leveraging different signal processing and feature extraction algorithms, coupled with statistical analysis and machine learning methods, many systems can realize activity recognition based on the CFR power~\cite{liu2019wireless}. Unlike them, in this work transmitter sends semantic features related to activity to the receiver. Therefore, we define the three dimensional (3D) semantic space, in which the activity recognition is realized by directly classifying the received semantic features, without recovering CFR power. Specifically, 
    the three dimensions of the defined semantic space are
    \begin{align}\label{fml12}
    \left\{ \begin{array}{l}
    x = frequency/\# \left( {bases} \right)\\
    y = {\rm{In}}\left( {amplitude/\# \left( {bases} \right)} \right)\\
    z = \# \left( {bases} \right)
    \end{array} \right.
    \end{align}
    where $\# (basis)$ is the number of semantic bases employed for semantic encoding\footnote{According to previous derivation, one can see that the initial phase of $H_D$ are determined by ${d_{{l_d}}}\left( 0 \right)$ and ${\varphi _{s{l_d}}}$, which contain no activity-related information. Therefore, the initial phase is not considered in the construction of the semantic space.}. Taking the signal in Fig.~\ref{exp ecd} as an example, we map the semantic features into 3D semantic space and obtain the averages corresponding to both activities, the results are shown in Fig.~\ref{MP exp}. It is clear that the bases of different activities are distributed in different parts of the 3D semantic space, creating favorable conditions for realizing recognition in the semantic space. This can be interpreted by the fact that different activities trigger different numbers of reflections with varying amplitude and frequency. We collect more sets of data for falling, walking, and sitting down in real-world scenarios and map bases into the 3D semantic space. The distribution of semantic features and corresponding averages, in 3D semantic space, are shown in Fig. ~\ref{PITS} and Fig. ~\ref{PITSAVRG}, respectively, which further verifies the feasibility of achieving activity recognition in the 3D semantic space.
    
    
    According to this observation, the kNN classifier is applied to classify semantic features in the semantic space. This process consists of two steps: offline training data collection and online recognition. During the offline phase, first, multiple sets of training data corresponding to each activity are collected to extract semantic features. Then, the features are mapped into the semantic space to obtain multiple averages for each activity. Here, the obtained averages are denoted as
    \begin{equation}
    {\bf{T}}{\rm{ = }}\left\{ {{T_1}, \ldots ,{T_m}, \ldots ,{T_M}} \right\},
    \end{equation}
    where ${T_m} = \left( {{x_m},{\rm{ }}{y_m},{\rm{ }}{z_m},{\rm{ }}{l_m}} \right)$, ${x_m}$, ${y_m}$, and ${z_m}$ are the coordinates of averages in semantic space, ${l_m} \in \left\{ {{A_1}, \ldots ,{A_K}} \right\}$ is the label. In this way, we can construct a cluster containing multiple labeled averages for each activity in 3D semantic space. In the online classification process, the test average of the to-be-classified data in the 3D semantic space is first obtained. After that, $K$ training averages closest to the test average in the labeled averages are identified and the neighborhood covering the $K$ training averages is determined as ${N_k}$. Finally, the activity recognition can be realized via 
    \begin{equation}
    l = \mathop {\arg \max }\limits_{{A_k}} \sum\limits_{{T_m} \in {N_k}} {\mathbb{I}\left( {{l_m} = {A_k}} \right)},
    \end{equation}
    where $\mathbb{I}\left(  \cdot  \right)$ is the indicator function and $k = 1,{\rm{ }}2,{\rm{ }} \cdots ,{\rm{ }}K$.
    
\begin{figure}[t]
    \centering
    \includegraphics[height=3.5cm]{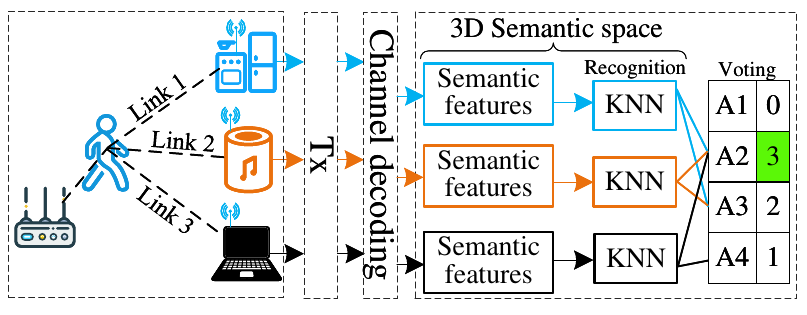} 
    \caption{The fusion of recognition results from multiple links.} 
    \label{links} 
    \end{figure}

    {\color{black} Considering the limited sensing capacity of a single transmission link and the possibility of numerous smart devices participating in physical world sensing, we combine the recognition results from multiple links via voting to further boost the performance.} As Fig.~\ref{links} illustrates, each link collects and processes the sensing data independently and transmits it to receiver for recognition and voting. Thereafter, the one with the largest number of votes is selected as the final recognition result\footnote{Due to transmission noise, packet loss or other causes, multiple activities may receive the same number of votes. For this case, we have two solutions. First, among the activities with the same number of votes, if one of them is the same as the previous recognition result, then this one is picked as the output. Otherwise, we randomly select one from them as the output.} and fed to MSP for further processing. 
    The activity recognition process in 3D semantic space is summarized in Algorithm 2.
    
    \begin{algorithm}[t]
    \caption{Activity recognition in semantic space.} 
    \label{Fills}
    \hspace*{0.02in} {\bf Input:}The labeled training data, test data, $K$ \\
    \hspace*{0.03in}{\bf Output:} The activity recognition result.
    \begin{algorithmic}[1]
    \For{each transmission link}
    \For{$m$=1:$M$}
    \State Calculating the distance between the test data and ${T_m}$ in semantic space
    \EndFor
    \State Sort the obtained distances in an ascending order
    \State Determine ${N_k}$ based on $K$ and sorted distances
    \State Activity recognition of each link: $l = \mathop {\arg \max }\limits_{{A_k}} \sum\limits_{{T_m} \in {N_k}} {\mathbb{I}\left( {{l_m} = {A_k}} \right)}$
    \EndFor
    \State Voting to obtain the final recognition result
    \end{algorithmic}
    \end{algorithm}
    
    \subsection{Contest Theory-based Sensing Data Market}
    {\color{black}To further enhance the user experience in Metaverse, we design an effective incentive mechanism for the sensing data market to promote transmitters uploading data more frequently. Concretely, under an available bonus amount, our goal is not to promote a specific transmitter to achieve the best possible performance or allocate network resource fairly, but to encourage all transmitters to redouble their efforts. Thus, unlike auction theory, contract theory and game theory, which are already widely used in wireless communication networks, a suitable solution here is to use the bonus to host a contest among transmitters. Here, the contest is defined as a game, in which contestants, i.e., the transmitters, must exert irreversible effort to win awards, which are determined according to the quality of the outputs.} In this section, we use the contest theory to examine the payoff of transmitters and the payment of receiver~\cite{corchon2018contest}.  
    \subsubsection{Effort, Capability, and Awards}
    In the context of wireless sensing in Metaverse, contestants are transmitters and their efforts can be regarded as the frequencies of uploading sensing data, i.e., let $f_n$ denote the frequency that the $n^{\rm th}$ transmitter sends the sensing data to the receiver for activity recognition in a time period $T$. Thus, the cost function for the $n^{\rm th}$ transmitter can be expressed as a twice differentiable function, i.e., ${{\cal C}_n}\left( {{a_n},{f_n}} \right)$, that satisfies~\cite{corchon2018contest}
    \begin{equation}\label{sasdf}
    \left\{ \begin{gathered}
    \frac{{\partial {C_n}\left( {{a_n},{f_n}} \right)}}{{\partial {f_n}}} > 0, \hfill \\
    \frac{{\partial {C_n}\left( {{a_n},{f_n}} \right)}}{{\partial {a_n}}} < 0, \hfill \\
    \frac{{{\partial ^2}{C_n}\left( {{a_n},{f_n}} \right)}}{{\partial {a_n}\partial {f_n}}} < 0, \hfill \\ 
    \end{gathered}  \right.
    \end{equation}
    where ${a_n}$ denotes the capability of the $n^{\rm th}$ transmitter. The last inequality in \eqref{sasdf} implies that the more capable a contestant is, the easier it is to increase the output. 
    
    The capability of transmitter can be measured in terms of the time that it takes to perform each upload. Specifically, a more capable transmitter can achieve faster data uploads because of better channel conditions and faster processing speed. Let $T_P$ and $T_P^{\left( S \right)}$ denote the time required for receiver to perform activity recognition using the original sensing data and semantic basis, respectively, and $T_S$ represents the time required to obtain the semantic bases from the sensing data, and $a_n$ can be expressed as
    \begin{align}\label{ageg2}
    a_n ={\left( {\frac{{{\cal S}\left( D \right)}}{{{C_n}}} + {T_S} + T_P^{\left( S \right)}} \right)^{ - 1}} 
    = {\left( {\frac{{{D_S}}}{{{C_n}}} + {T_S} + T_P^{\left( S \right)}} \right)^{ - 1}},
    \end{align}
    where $S$ is the semantic-aware encoding operator, and $S\left( D \right)$ is the bit amount of semantic bases for one wireless sensing. 
    
    For a given upload frequency, a more capable transmitter can upload faster, resulting in a smaller cost. Taking \eqref{sasdf} into consideration, the cost function can be denoted as ${{\cal C}_n}\left( {{a_n},{f_n}} \right) ={f_n}/{a_n}$. Sorting {\small $N_T$} transmitters in descending order according to efforts to obtain $ \left\{ {{f_1}, \ldots ,{f_{{N_T}}}} \right\} $, then the award that the $n^{\rm th}$ transmitter receives can be denoted as $r_n$ $\left(1 \le n \le {N_A}\le{N_T}\right)$, where ${N_A}$ is the number of awards, $ {r_1} \ge {r_2} \ge  \cdots  \ge {r_{{N_A}}} $, and $r_n = 0$ when $n>N_A$. On this basis, all the awards are summed up to obtain the total amount that receiver pays, which is expressed as $ {r_T} = \sum\limits_{n = 1}^{{N_A}} {{r_n}}$.
    
    \subsubsection{Optimal Effort Analysis}
    The utility of the $n^{\rm th}$ transmitter participating in the contest can be expressed as
    \begin{align}
    &\pi \left( {{a_n},{f_n}} \right) 
    \notag\\
    &=\! \left\{\!\!\! {\begin{array}{*{20}{l}}
    {{r_m} - {{\cal C}_n}\left( {{a_n},{f_n}} \right)}, \: {\textrm{if the transmitter is ranked in $m^{\rm th}$,}}\\
    { - {{\cal C}_n}\left( {{a_n},{f_n}} \right)}, \:\:{\textrm{if the transmitter is not awarded.}}
    \end{array}} \right.
    \end{align}
    Then, the expectation of the utility can be derived as follows:
    \begin{equation}\label{eall}
    \mathbb{E}\left[ {\pi \left( {{a_n},{f_n}} \right)} \right] = {\cal R}\left( {{f_n}} \right) - {{\cal C}_n}\left( {{a_n},{f_n}} \right),
    \end{equation}
    where ${\cal R}\left( {{f_n}} \right)$ denotes the expectation of the award that the $n^{\rm th}$ transmitter receives. Considering that all transmitters are rational, the following lemma is first derived
    \begin{lemma}\label{lemma1}
    The more capable the transmitter is, the more effort is exerted in the contest. In other words, a transmitter with a higher transmission rate and a shorter activity recognition processing time uploads sensor data more frequently.
    \end{lemma}
    \begin{IEEEproof}
    Please refer to Appendix~\ref{lemma1a}.
    \end{IEEEproof}
    
    We consider that the CDF of capability in the population is represented by a continuous function $\mathcal{G}\left( a \right)$, and each transmitter knows its own capability, but has only probabilistic estimates of the capabilities of fellow potential contestants, i.e., other transmitters. Assuming the transmitter with capability $a_n$ chooses effort $f_n$ and its remuneration consists solely of its prize, so that a transmitter with capability $a_n$ chooses to participate in the contest only if its surplus, $\mathcal{R}\left(a_n, y\left(a_n\right)\right)$, is no less than the opportunity cost of participation.
    
    According to Lemma~\ref{lemma1}, a transmitter with greater capability will choose a higher level of effort. Hence, the probability that a transmitter with capability $a$ obtains the $j^{\rm th}$ highest prize is simply the probability that it has the $j^{\rm th}$ highest capability among the $N_T$ potential transmitters. On this basis, the expected award won by the $n^{\rm th}$ transmitter with capability $a_n$ is
    \begin{equation}\label{eawa}
    {\cal R}\left( {{a_n}} \right) = \sum\limits_{m = 1}^{{N_A}}{u\left( r_m \right)} {\left(\!\!\! {\begin{array}{*{20}{c}}
    {{N_T} - 1}\\
    {m - 1}
    \end{array}} \!\!\!\right)} {{\cal G}^{{N_T} - m}}\left( {{a_n}} \right){\left( {1 - {\cal G}\left( {{a_n}} \right)} \right)^{m - 1}},
    \end{equation}
    where $u\left( r_m \right)$ is the expected utility of the $m^{\rm th}$ award decided by the risk preference~\cite{tremewan2020behavioral}. Specifically, $u\left( r_m \right) = r_m $ denotes risk neutral and $u\left( r_m \right) = \ln \left( r_m \right)$ means risk averse. 
    
    A transmitter does not know the capabilities of other transmitters, but it will estimate the distribution of the capabilities, i.e., ${{\cal G}_{X}}\left( {x} \right)$. Recall that capability is a function of data rate, semantic encoding time, and activity recognition time, here the transmitter will estimate its expected utility in the contest by assuming that the data rate of other transmitters obeys a uniform distribution\footnote{Here the uniform distribution is considered and, in fact, the proposed analysis method can be easily extended to other distributions.}. To obtain a close-form expression of ${\cal R}\left( {{a_n}} \right)$, the ${\cal G}\left( {{a_n}} \right)$ is analyzed first.
    \begin{lemma}\label{falkjl2}
    From the $n^{\rm th}$ transmitter's point of view, the probability that the capabilities of other transmitters is less than $a_n$, i.e., ${\cal G}\left( {{a_n}} \right)$, is
    \begin{align}\label{afe1235r}
    {{\cal G}}\left( {a_n} \right)
    = \left\{ {\begin{array}{*{20}{l}}
    {\frac{1}{{\Delta}}\frac{{{a_n}{D_S}}}{{1 - {a_n}{T_S} - {a_n}T_P^{\left( S \right)}}},\quad {\textrm{Condition 1}} } \\ 
    {0,\quad {\textrm{otherwise}},}
    \end{array}} \right.
    \end{align}
    where {\textrm{Condition 1}} is 
    \begin{equation}
    {0 \!< {a_n} <\! {{\left( {\frac{{{D_S}}}{{{\Delta}}}\! + {T_S} + T_P^{\left( S \right)}} \right)}^{ - 1}}}.
    \end{equation}
    \end{lemma}
    \begin{IEEEproof}
    Please refer to Appendix~\ref{falkjl2a}.
    \end{IEEEproof}
    
    With the Lemma~\ref{falkjl2}, the optimal effort for each transmitter is obtained as in Theorem 1.
    \begin{them}\label{ageklgj}
    The optimal effort of each transmitter is given as
    \begin{equation}\label{skjf}
    f_n^* = {a_n}R\left( {{a_n}} \right) - \sum\limits_{m = 1}^{{N_A}} {\frac{{u\left( r_m \right){a_n}{H_f}}}{{\Gamma\!\left( {{N_T} - 1} \right)}}} \left(\!\!\! {\begin{array}{*{20}{c}}
    {{N_T} - 1} \\ 
    {m - 1} 
    \end{array}} \!\!\!\right)\!{\left( {\frac{{{D_S}}}{\Delta }} \right)^{{N_T} - m}},
    \end{equation}
    where $H_f$ is derived as \eqref{fakhj}, shown at the bottom of the next page.
    \newcounter{mycount2}
    \begin{figure*}[b]
    \normalsize
    \setcounter{mycount2}{\value{equation}}
    \hrulefill
    \vspace*{4pt}
    {\small \begin{equation}\label{fakhj}
    {H_f}{\text{ = }}H_{1,1:3,3;1,1}^{0,1:1,2;1,1}\!\!\left(\!\! \!\!{\left. {\begin{array}{*{20}{c}}
    {{a_n}\left( {{T_S} + T_P^{\left( S \right)} + \frac{{{D_S}}}{\Delta }} \right)} \\ 
    {\frac{{\Delta {T_S} + \Delta T_P^{\left( S \right)}}}{{ - \Delta }}{a_n}} 
    \end{array}} \right|\begin{array}{*{20}{c}}
    {\left( {m - {N_T};1,1} \right):\left( {0,1} \right)\left( {m - {N_T} - 1;1} \right)\left( {m,1} \right);\left( {2 - {N_T},1} \right)} \\ 
    {\left( {m - {N_T} - 1;1,1} \right):\left( {0,1} \right)\left( { - 1,1} \right)\left( {m - {N_T},1} \right);\left( {0,1} \right)} 
    \end{array}} \!\!\!\right)
    \end{equation}}
    \setcounter{equation}{\value{mycount2}}
    \end{figure*}
    \addtocounter{equation}{1}
    \end{them}
    \begin{IEEEproof}
    Please refer to Appendix~\ref{ageklgja}.
    \end{IEEEproof}
    
    \subsubsection{Optimal Awards Analysis}
    The receiver adjusts the awards to maximize the sum of transmitter efforts by solving 
    \begin{equation}\label{pro1}
    \begin{array}{*{20}{l}}
    {\mathop {\max }\limits_{N_A,{r_1}, \ldots ,{r_{{N_A}}}} }&{\sum\limits_{n = 1}^{{N_T}} {f_n^*} } \\ 
    {\:\qquad{\text{s}}{\text{.t}}{\text{.}}}&{\sum\limits_{m = 1}^{{N_A}} {{r_m}}  \leqslant r}.
    \end{array}
    \end{equation}
    With the help of Theorem \ref{ageklgj}, we can rewrite $f_n^*$ as
    \begin{equation}
    f_n^* = \sum\limits_{m = 1}^{{N_A}} {u\left( r_m \right)\mathcal{F}_n\left( {{a_n},{N_T},m,{D_S},{T_S},T_P^{\left( S \right)},\Delta } \right)} .
    \end{equation}
    \begin{them}\label{them2}
    By solving the optimization problem \eqref{pro1}, we can obtain the optimal award setting as $r_1 = r$ and $r_2=\cdots=r_{N_T}=0$, when the transmitters are risk neutral.
    \end{them}
    \begin{IEEEproof}
    Please refer to Appendix~\ref{them2a}
    \end{IEEEproof}
    
    \begin{them}\label{them3}
    If the transmitters are risk averse, we can obtain the optimal award setting as follows:
    \begin{equation}\label{afeaf}
    {r_\ell } = \frac{{r\sum\limits_{n = 1}^{{N_T}} {{\cal F}\left( {{a_n},{N_T},\ell ,{D_S},{T_S},T_P^{\left( S \right)},\Delta } \right)} }}{{\sum\limits_{m = 1}^{{N_A}} {\sum\limits_{n = 1}^{{N_T}} {{\cal F}\left( {{a_n},{N_T},m,{D_S},{T_S},T_P^{\left( S \right)},\Delta } \right)} } }},
    \end{equation}
    where $\ell = 1,\ldots,N_A$.
    \end{them}
    \begin{IEEEproof}
    Please refer to Appendix~\ref{them3a}.
    \end{IEEEproof}
    
    

    \section{Numerical Analysis}\label{fealjk}
    In this section, we first introduce the experimental configurations. Then, based on the collected data, the effectiveness of activity recognition in semantic space is verified. After that, the impact of wireless transmission on recognition accuracy is simulated. Finally, the effect of semantic transmission and incentive mechanisms on the overall system is investigated.
    
    \subsection{Experimental Platform, Scenario, and Data Collection}
\begin{figure}[t]
    \centering
    \includegraphics[height=3.8cm]{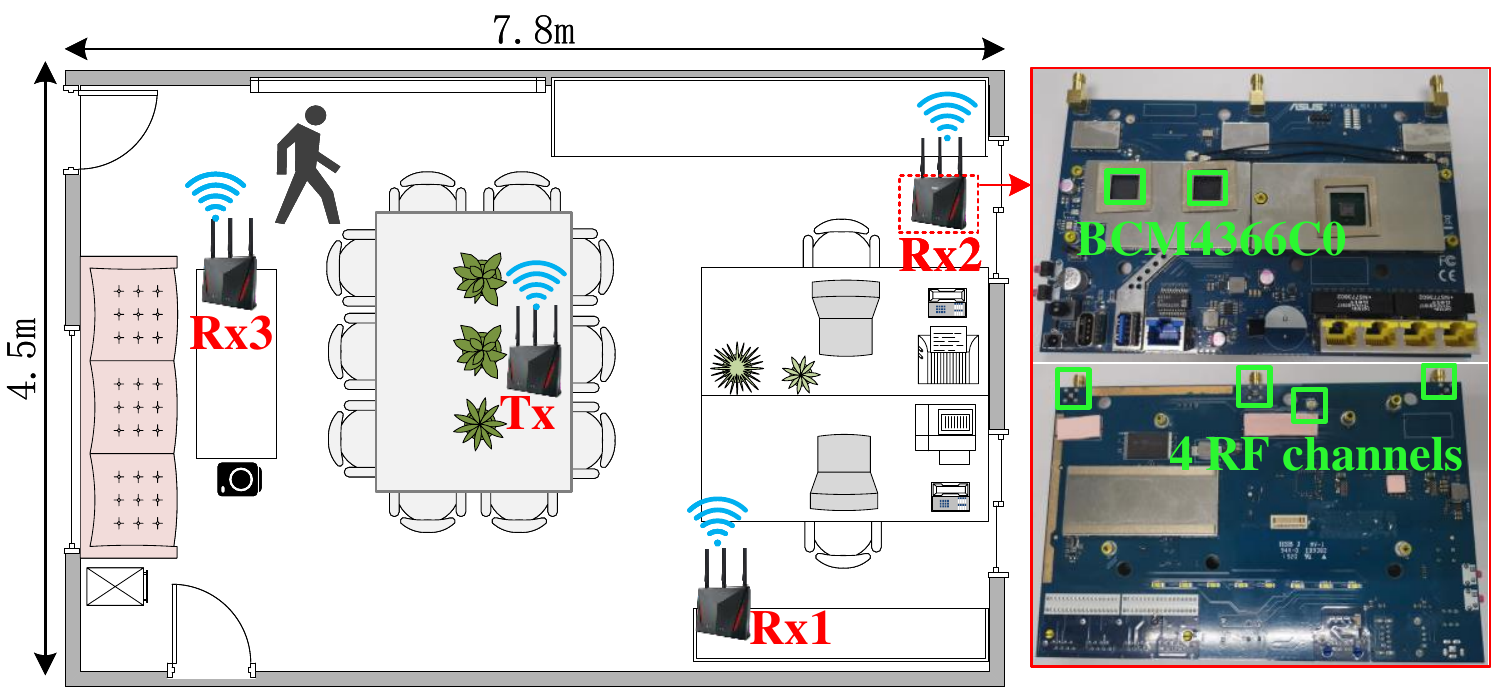} 
    \caption{The test scenario and hardware equipment.} 
    \label{hardware} 
    \end{figure}

    An IEEE 802.11ac protocol based experimental platform is built to collect CSI in real-world scenarios to validate the activity recognition in semantic space. The specific hardware and test scenario are shown in Fig. ~\ref{hardware}. The test scenario is a 7.8m × 4.5m office room with different furniture, such as conference table, chairs, computers, and bookcases. Inside the office room, we placed a transmitter and 3 receivers, denoted as Tx and Rx1-Rx3, to transmit and receive wireless signals, respectively. The experimental equipment is an access point (AP) equipped with a BCM4366C0 chip and the Nexmon toolbox~\cite{gringoli2019free}, as shown on the right side of Fig.~\ref{hardware}. During the data collection process, the frequency of the signal is set to 5.805 GHz, the signal bandwidth is 80 MHz, and the default transmission rate of the CSI data packet is 600 Hz. While each AP has multiple antennas, the Tx and Rx are controlled to operate via one antenna.
    \subsection{Performance Evaluation }
    
\begin{figure*}[htbp]
    \centering
    \subfigure[]{
    \begin{minipage}[t]{0.5\linewidth}
    \centering
    \includegraphics[width=10.5cm]{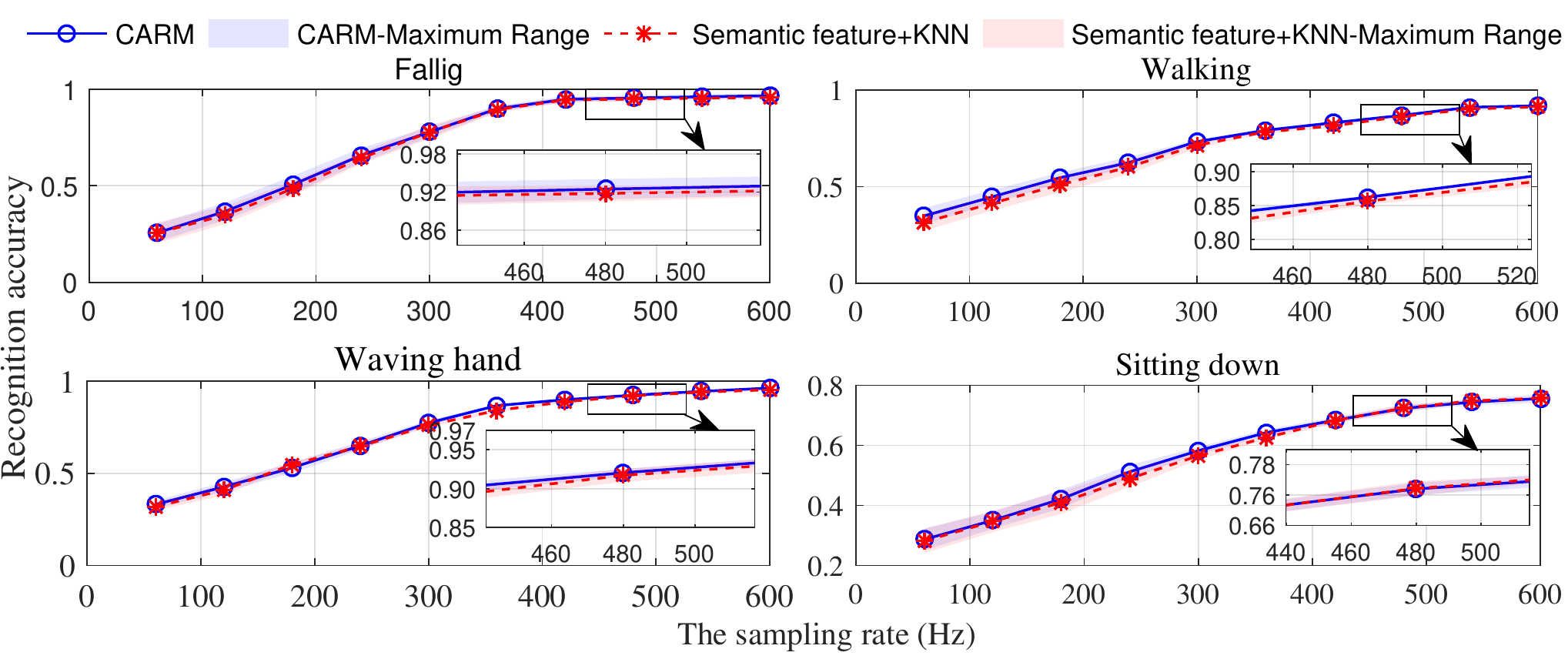}
    \end{minipage}%
    }%
    \subfigure[ ]{
    \begin{minipage}[t]{0.65\linewidth}
    \centering
    \includegraphics[width=5.7cm]{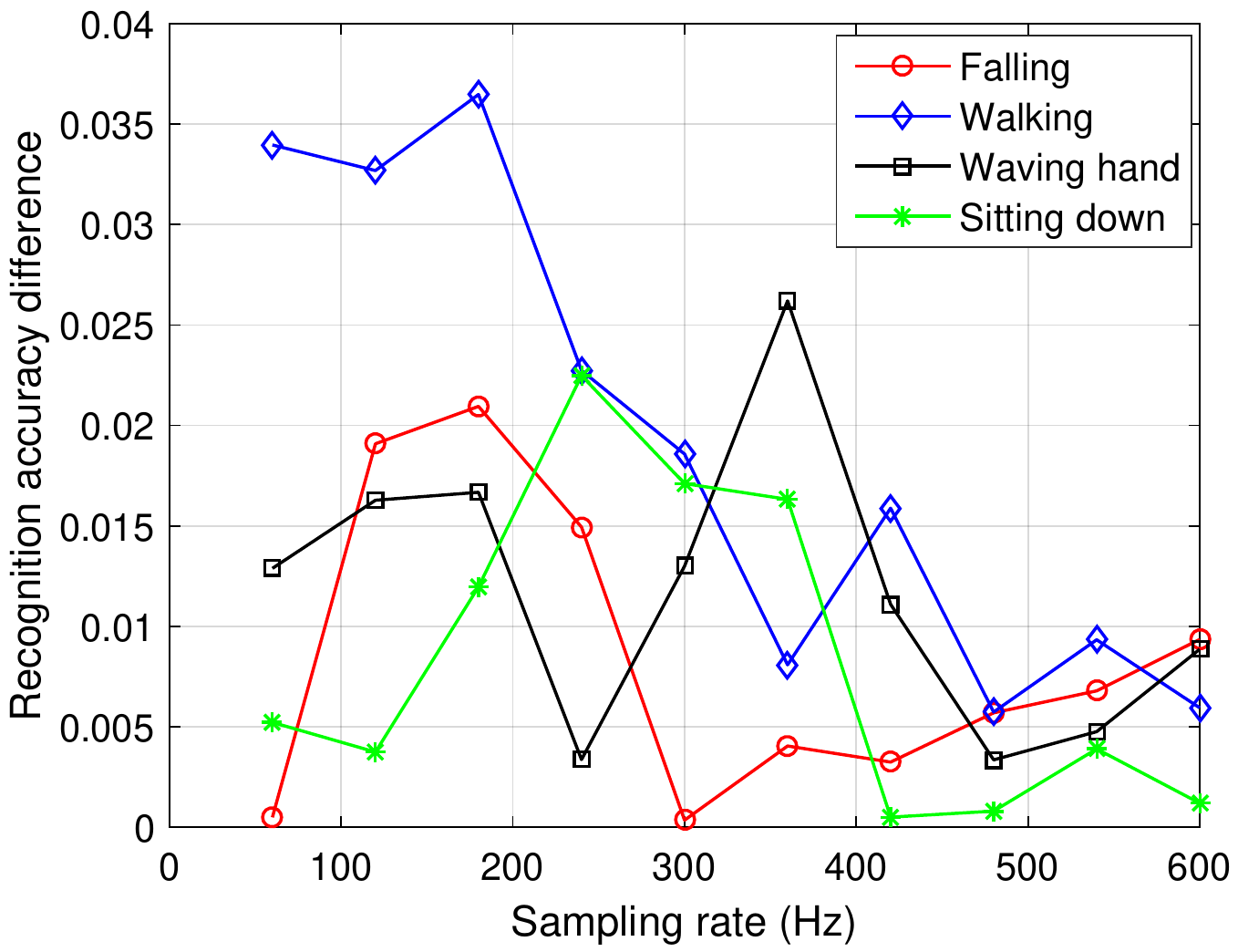}
    \end{minipage}
    }%
    \centering
    \caption{{\color{black}The activity recognition accuracy in the 3D semantic space.} }
    \label{Acrc} 
    \end{figure*}

    {\textbf{The activity recognition in semantic space.}} {\color{black}We collected measurement data of four activities and evaluated the recognition performance in semantic space. To demonstrate the effectiveness of our framework, we compare it with CRAM~\cite{wang2017device} and the results are shown in Fig.~\ref{Acrc}. First, the Fig.~\ref{Acrc} (a) shows the performance of the both systems improves with the increase of the sampling rate and the general trend is nearly identical, as presented by the partially enlarged result.} The reason is that the reduce of packet transmission rate makes the obtained original CSI unable to fully describe the high-frequency features, affecting the recognition performance. Second, the recognition accuracy in semantic space is close to that of CRAM. {\color{black}As an example, Fig.~\ref{Acrc} (b) indicates that for walking, the maximum recognition accuracy difference between two systems is 0.0365, and the minimum difference is 0.0057.} For other activities, the difference in recognition accuracy is smaller. The results reveal that the performance of activity recognition in semantic space is comparable to that of CRAM, verifying the effectiveness of semantic encoding, as well as activity recognition in semantic space.

\begin{figure*}[htbp]
    \centering
    \subfigure[]{
    \begin{minipage}[t]{0.32\linewidth}
    \centering
    \includegraphics[width=5.5cm]{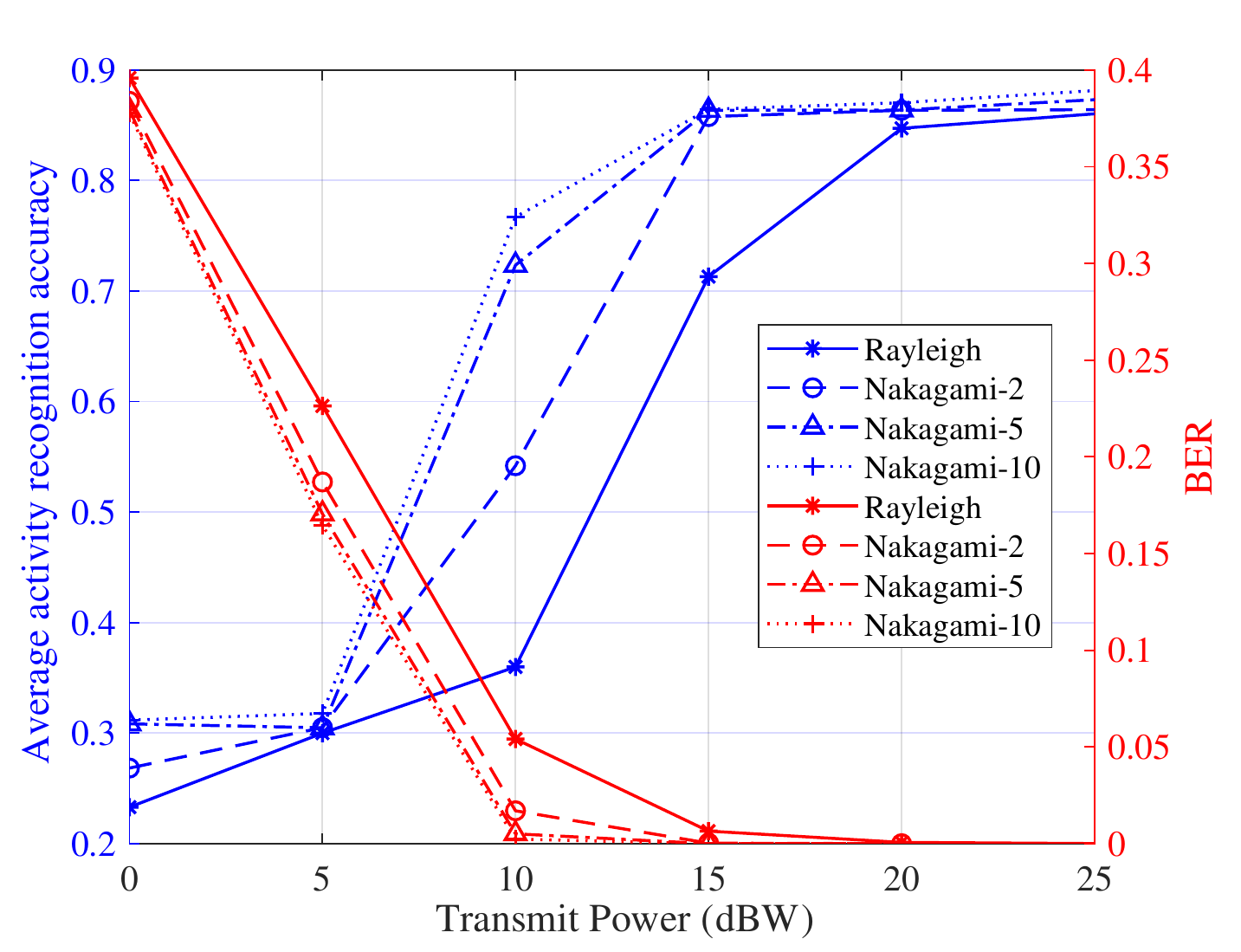}
    \end{minipage}%
    }%
    \subfigure[ ]{
    \begin{minipage}[t]{0.32\linewidth}
    \centering
    \includegraphics[width=5.5cm]{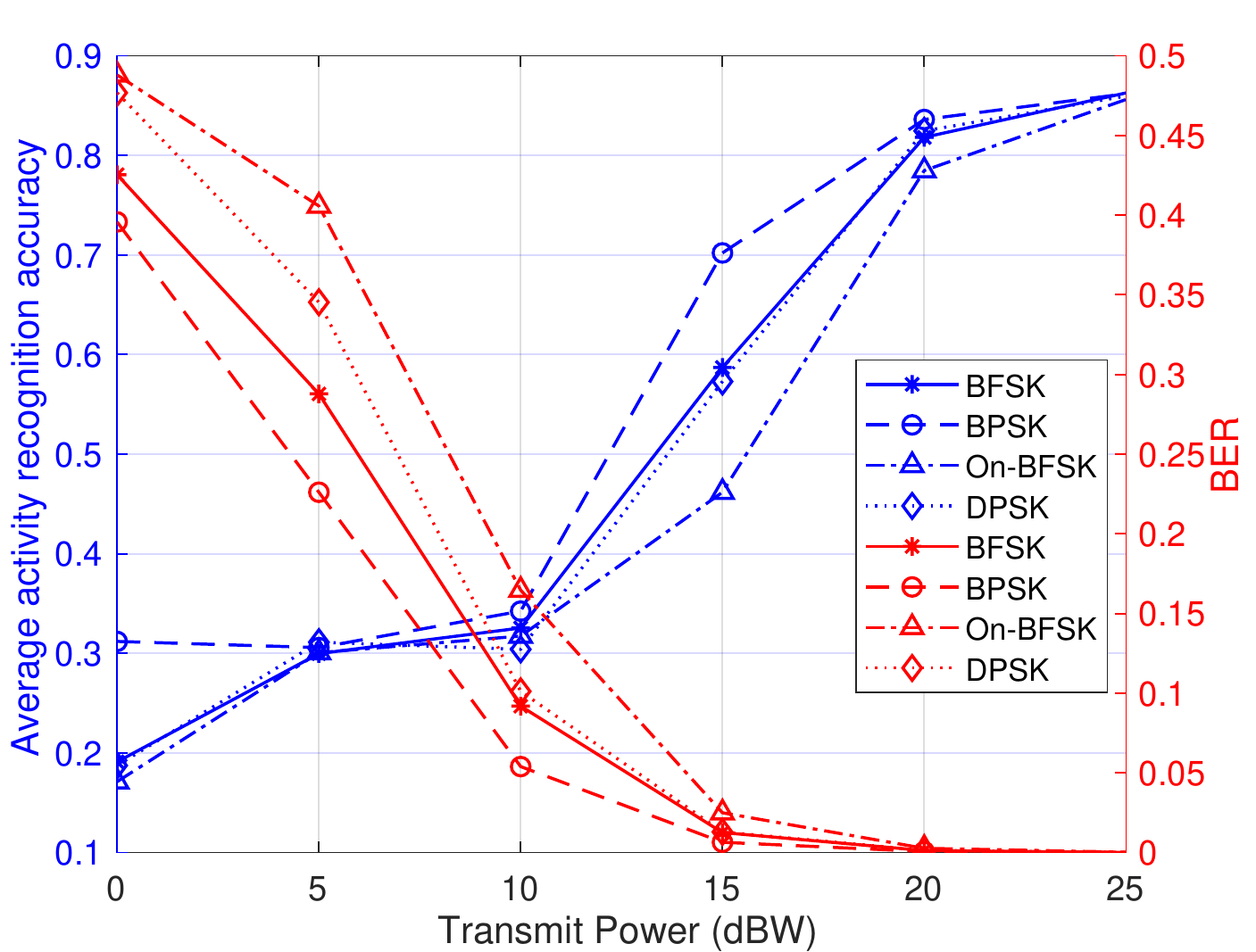}
    \end{minipage}
    }%
     \subfigure[ ]{
    \begin{minipage}[t]{0.33\linewidth}
    \centering
    \includegraphics[width=4.8cm]{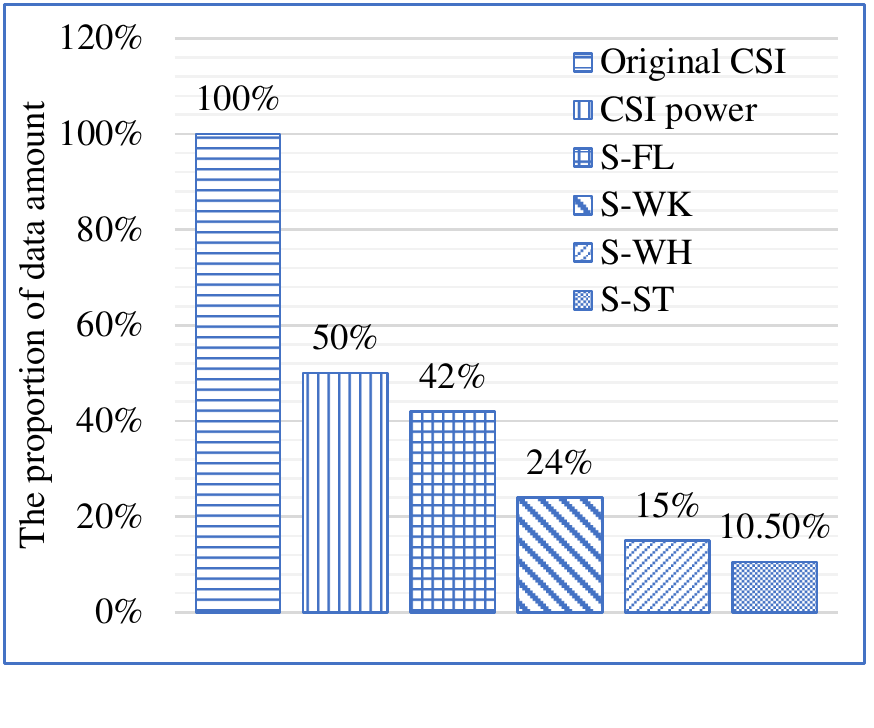}
    \end{minipage}
    }%
    \centering
    \caption{{\color{black}The impact of wireless transmission on recognition accuracy and the data amount comparison before and after semantic encoding.} }
    \label{FDING} 
    \end{figure*}

    {\textbf{The impact of wireless transmission on recognition accuracy.}} We simulated the wireless transmission of the semantic feature to analyze the impact of transmission on activity recognition. Figure~\ref{FDING} (a) gives the activity recognition accuracy in semantic space under different small-scale fading conditions obtained using the $H$-fading model in Section~\ref{fading}. One can see that the BER of Rayleigh is larger than the Nakagami for different transmit power. Meanwhile, for Nakagami, increasing the order can reduce the BER. Therefore, as denoted by black lines, the Nakagami-10 has the highest recognition accuracy, followed by Nakagami-5, then Nakagami-2, and the lowest is Rayleigh. The reason is that smaller order in the Nakagami fading means stronger multipath effect. In addition, increasing the transmit power can reduce the BER, thereby improving the recognition accuracy. When the transmit power reaches 25 dBw, the recognition accuracy of Nakagami-10 is 0.8831, which is comparable to 0.8953 of recognition in semantic space without transmission and 0.9010 of CRAM, further demonstrating the feasibility of semantic encoding and activity recognition in semantic space. 
    
    Besides small-scale fading, the impact of modulation methods on the recognition accuracy is analyzed under the Rayleigh fading condition, and the results are shown in Fig.~\ref{FDING} (b). With the same transmit power, the BPSK has the lowest BER, then BFSK, followed by DPSK, and orthogonal non-coherent BFSK (ON- BFSK) scheme has the largest BER. As a result, BPSK has the best activity recognition performance, while ON-BFSK performs the worst. Meanwhile, similar to the result in Fig.~\ref{FDING} (a), the recognition accuracy improves with the increase of transmit power. When transmit power arrives at 25 dBw, the recognition accuracies corresponding to ON-BFSK and BPSK are 0.8560 and 0.8618, respectively, which are believed to be acceptable.

    {\textbf{Data amount comparison before and after semantic encoding.}} Let the transmitter upload data 12 times per second\footnote{According to \cite{vasisht2016decimeter}, the channel coherence time in an indoor environment is around 84 ms and  $600 \times 0.084 \approx 50$. Therefore, with a packet transmission rate of 600 Hz, we can assume that the transmitter uploads data 12 times per second, i.e., uploads every 50 data packets are collected.}, {\color{black}then the data amount of the original CSI, CSI power, and semantic encoded data is illustrated in Fig.~\ref{FDING}. (c), where, for example, S-FL refers to the semantic encoding of falling data. As can be seen, the average amount of data uploaded each time after semantic encoding is approximately $27.87\%$ of that before encoding, demonstrating the proposed semantic encoding algorithm can greatly reduce the amount of data while ensuring the system performance. This creates excellent space for further establishing the incentive mechanism to promote data uploading. Besides that, the amount of encoded data corresponding to different activities is different. This is understandable, since the part of the body which caused main reflections is different for various activities. For instance, during walking the arms and legs reflect a large number of signals in addition to the torso, while for sitting down the reflections are mainly caused by torso. Therefore, different numbers of semantic bases are needed to encode data corresponding to various activities, leading to differences in the amount of encoded data.}

    {\small \begin{table}
    \caption{Network Parameters for $3$ transmitters.}
    \label{grg}
    \centering
    \renewcommand{\arraystretch}{1}
    {\small\begin{tabular}{|m{4cm}<{\centering}|m{1cm}<{\centering}|m{1cm}<{\centering}|m{1cm}<{\centering}|}
    \toprule
    \hline
    \textbf{Parameters} & $1_{\rm st}$ transmitter & $2_{\rm nd}$ transmitter& $3_{\rm rd}$ transmitter \\
    \hline
    Time to obtain the semantic basis, $T_S$ & \multicolumn{3}{c|}{9.797 ms} \\
    \hline
    Time to perform activity recognition, $M_U$ & \multicolumn{3}{c|}{5 ms} \\
    \hline
    Estimated maximum data rate, $\Delta$ & \multicolumn{3}{c|}{8 Mbps} \\
    \hline
    Real data rate, ${C_n}$ & {7 Mbps}
    & {6 Mbps} & {5 Mbps} \\
    \hline
    Number of bits generated in once sensing, $D$ & \multicolumn{3}{c|}{96000}\\
    \hline
    Number of bits of semantic bases in once sensing, $D_S$ & \multicolumn{3}{c|}{7200}\\
    \hline
    \bottomrule
    \end{tabular}}
    \end{table}}
    \begin{figure*}
    \centering
    \includegraphics[width=0.95\textwidth]{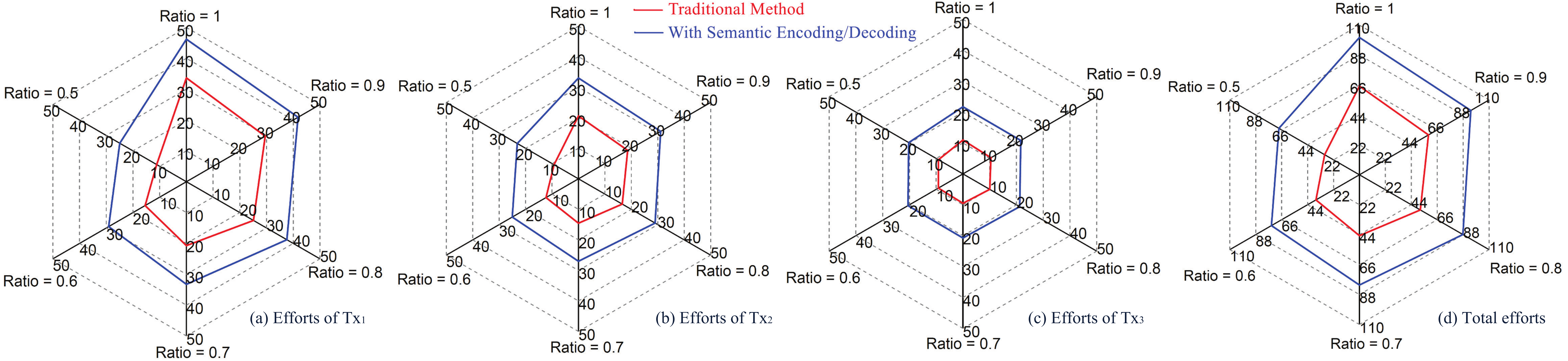}
    \caption{The efforts of three risk neutral transmitters when the semantic methods are used or not, under six different award setting schemes.}
    \label{zhizhu}
    \end{figure*}
    {\textbf{The effects of the semantic method and the award setting schemes.}}  {\color{black} We consider the case where there are three risk neutral transmitters and two awards are set in the contest. For a comprehensive analysis, six award setting schemes are compared, i.e., the ratios of the first award and total award are $0.5$, $\ldots$, $1$. Besides that, some other parameter configurations are shown in Table~\ref{grg}, where the processing time and the number of bits are obtained from real-world experiments, and the data rate values are typical in wireless networks~\cite{yadav2021review}. } Figure~\ref{zhizhu} (a)-(c) shows the efforts of three transmitters when the semantic encoding is used or not. From the results, one can observe that the use of semantic encoding motivates transmitters to upload sensing data under all award schemes in the contest. The reason is that the data amount of semantic features is much smaller than that of the raw sensing data, making the transmitter using semantic encoding has lower cost and higher capability. Therefore, under the same expected award, each transmitter is willing to put in more effort to win the contest. Another interesting insight from Fig.~\ref{zhizhu} (d) is that the sum of three transmitters' upload frequencies achieves the maximum when the total award is set as the only first prize. This result demonstrates our Theorem~\ref{them2}. {\color{black}Compared with the uniform allocation scheme, the optimal award allocation strategy can increase the sum of uploading frequency by $27.47\%$.}
    \begin{figure}[t]
    \centering
    \includegraphics[width = 0.4\textwidth]{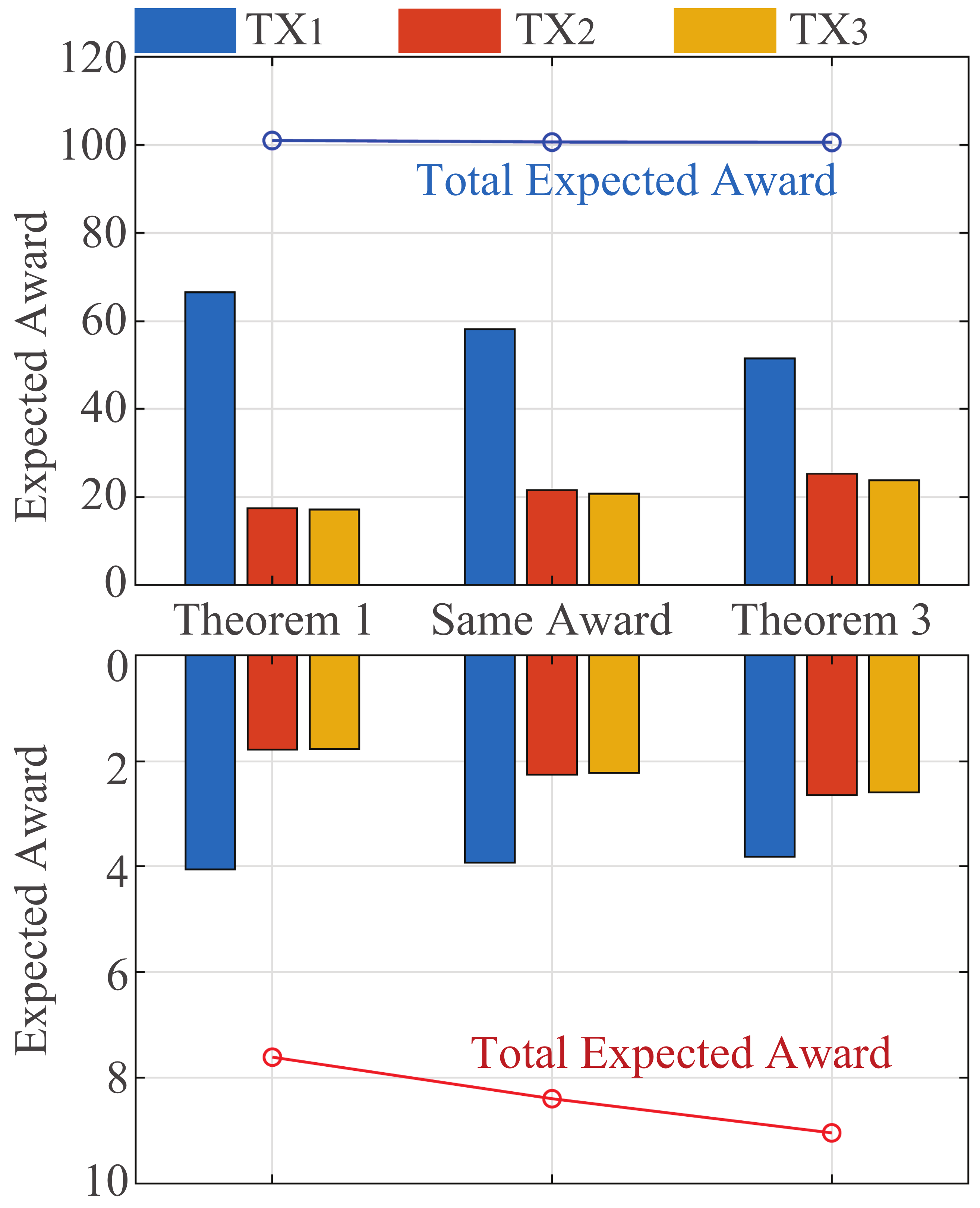} 
    \caption{The expected awards of transmitters under different award scheme settings and risk appetites.} 
    \label{risk}
    \end{figure}
    
    {\textbf{The effects of the risk appetite and the award setting.}} Since the optimal award scheme is related to the transmitters' risk appetite, here we further discuss the case where transmitter is risk averse. From \eqref{eall}, one can observe that the greater the expected award, the more willing transmitters are to put effort in the contest. In Fig.~\ref{risk}, we show the awards that transmitters expect to receive under different award scheme settings and risk appetites. Again, there are three transmitters and three awards. The parameters related to transmitters are shown in Table~\ref{grg}. {\color{black} Three different award setting schemes are considered, i.e., all awards were allocated to the first prize (Theorem~\ref{them2}), and the awards are divided equally by three prizes, and awards are allocated following Theorem~\ref{them3}. When transmitters are risk neutral, we can see that reducing the proportion of first prize awards results in a significant reduction, i.e., $22\%$, in the expected award of the most able transmitter. Although the expected awards for the remaining participants increase, the total expected award for the three transmitters decreases slightly. However, when transmitters are risk averse, our derived optimal award setting scheme, i.e., Theorem~\ref{them3}, brings only a slight decrease in the expected award for the most capable transmitter, i.e., $7\%$, compared to the optimal award setting scheme when the user is risk neutral. However, the expected awards of the rest of the participants increase obviously. Finally, the total expected award can be increased by $20\%$.}
    \section{Conclusion}
    A semantic aware transmission framework has been proposed in this paper, for transforming and transmitting sensing data from the physical world to MSP in Metaverse. Two breakthroughs lie in paper are semantically aware sensing data transmission and the contest theory based incentive mechanism, respectively. We have proposed the first semantic encoding algorithm for sensing data, reducing the amount of data significantly and lowering the storage and transmission costs, while ensuring the performance of MSP in Metaverse. {\color{black}In addition, we have established a contest-based incentive mechanism to boost all transmitters' data uploading frequency by setting rewards, which provides stronger support for MSP to enhance its QoS.} The results show that the average data amount after semantic encoding is only $27.87\%$ of the original sensing data and the data uploading frequency is increased by $27.47\%$ with the help of proposed incentive mechanism. For future work will be extended to semantic aware transmission of other sensing information.

    \begin{appendices}
    \section{Proof of Lemma~\ref{lemma1}}\label{lemma1a}
    \renewcommand{\theequation}{A-\arabic{equation}}
    \setcounter{equation}{0}
    The transmitter maximizes the expectation of utility by adjusting its effort. Thus, the optimal effort of the $n^{\rm th}$ transmitter, i.e., $ {f_n^*} $, satisfies the following conditions:
    \begin{equation}\label{con}
    \left\{ \begin{gathered}
    {\left. {\frac{{\partial \mathbb{E}\left[ {\pi \left( {{a_n},{f_n}} \right)} \right]}}{{\partial {f_n}}}} \right|_{{f_n} = f_n^*}} = 0 \hfill \\
    {\left. {\frac{{{\partial ^2}\mathbb{E}\left[ {\pi \left( {{a_n},{f_n}} \right)} \right]}}{{\partial f_n^2}}} \right|_{{f_n} = f_n^*}} < 0 \hfill \\ 
    \end{gathered}  \right.
    \end{equation}
    Substituting \eqref{eall} and \eqref{eawa} into \eqref{con}, we have $ {R\left( {{f_n}} \right)} $ satisfies 
    \begin{equation}\label{afe}
    \frac{{\partial R\left( {f_n^*} \right)}}{{\partial f_n^*}} = \frac{{\partial {C_n}\left( {{a_n},f_n^*} \right)}}{{\partial f_n^*}} = \frac{1}{{{a_n}}}
    \end{equation}
    and
    \begin{equation}\label{faeg}
    \frac{{{\partial ^2}R\left( {f_n^*} \right)}}{{\partial f{{_n^*}^2}}} < 0.
    \end{equation}
    Since ${f_n^*}$ is a function of $a_n$ and $r_n$, we take the partial derivative of \eqref{afe} with respect to ${f_n^*}$ to obtain $\frac{{{\partial ^2}R\left( {f_n^*} \right)}}{{\partial f{{_n^*}^2}}} = \frac{{ - 1}}{{a_n^2}}\frac{{\partial {a_n}}}{{\partial f_n^*}}$. With the help of \eqref{faeg}, we have $\frac{{\partial f_n^*}}{{\partial {a_n}}} > 0$, which means that, under the optimal effort strategy, the more capable transmitter exerts more effort. The proof is completed.
    
    \section{Proof of Lemma~\ref{falkjl2}}\label{falkjl2a}
    \renewcommand{\theequation}{B-\arabic{equation}}
    \setcounter{equation}{0}
    The CDF of $C_n$ can be expressed as
    \begin{equation}\label{feqgk1}
    {F_{Cn}}\left( x \right) = \left\{ {\begin{array}{*{20}{l}}
    {\frac{x}{{{\Delta} }},\left( {{\Delta} > x > 0} \right)} \\ 
    {0.} 
    \end{array}} \right.
    \end{equation}
    According to \eqref{ageg2}, we have ${C_n} = \frac{{{D_S}}}{{\frac{1}{{a_n}} - {T_S} - T_P^{\left( S \right)}}}$. Thus, the CDF of $a_n$ can be expressed as ${F_{a_n}}\left( y \right) = {F_{Cn}}\left( {\frac{{y{D_S}}}{{1 - y{T_S} - yT_P^{\left( S \right)}}}} \right)$. Substituting \eqref{feqgk1} into ${F_{a_n}}\left( y \right)$, we can obtain \eqref{afe1235r} to complete the proof.
    
    \section{Proof of Theorem~\ref{ageklgj}}\label{ageklgja}
    \renewcommand{\theequation}{C-\arabic{equation}}
    \setcounter{equation}{0}
    Taking the derivative of ${\partial R\left( {{a_n}} \right)}$ with respect to ${a_n}$, we obtain $\frac{{\partial R\left( {{a_n}} \right)}}{{\partial {a_n}}} = \frac{{\partial R\left( {{a_n}} \right)}}{{\partial f_n^*}}\frac{{\partial f_n^*}}{{\partial {a_n}}} = \frac{1}{{{a_n}}}\frac{{\partial f_n^*}}{{\partial {a_n}}}$. Therefore, the optimal effort can be expressed as
    \begin{equation}
    f_n^* = {a_n}R\left( {{a_n}} \right) - \underbrace {\int_0^{{a_n}} {R\left( y \right){\rm{d}}y} }_I.
    \end{equation}
    Substituting \eqref{eawa} and \eqref{afe1235r} into $I$, we have
    \begin{align}\label{afef341}
    &I = \sum\limits_{m = 1}^{{N_A}} {\frac{{u\left( r_m \right)}}{{{\Delta ^{{N_T} - 1}}}}} \left(\!\! {\begin{array}{*{20}{c}}
    {{N_T} - 1} \\ 
    {m - 1} 
    \end{array}} \!\!\right)
    \notag\\
    &\int_0^{{a_n}} {\frac{{{{\left( {y{D_S}} \right)}^{{N_T} - m}}{{\left( {\Delta \! - \!\left(\! {\Delta {T_S} \!+ \!\Delta T_P^{\left( S \right)} \!+\! {D_S}} \!\right)y} \right)}^{m - 1}}}}{{{{\left( {1 - \frac{{\Delta {T_S} + \Delta T_P^{\left( S \right)}}}{\Delta }y} \right)}^{{N_T} - 1}}}}{\text{d}}y}.
    \end{align}
    Note that the integral in \eqref{afef341} is difficult to solve directly. To obtain the close-form solution of $I$, we perform an $H$-transformation~\cite[eq. (2.9.6)]{kilbas2004h} of the numerator as follows:
    \begin{align}\label{1243ehk}
    & \quad {\left( {y{D_S}} \right)^{{N_T} - m}}{\left( {\Delta  - \left( {\Delta {T_S} + \Delta T_P^{\left( S \right)} + {D_S}} \right)y} \right)^{m - 1}}
    \notag\\&
    = {D_S}^{{N_T} - m}{\Delta ^{m - 1}}{\left( {{T_S} + T_P^{\left( S \right)} + \frac{{{D_S}}}{\Delta }} \right)^{m - {N_T}}}
    \notag\\
    &\times 
    H_{1,1}^{1,0}\left( \!{\left. {\left( {{T_S} + T_P^{\left( S \right)} + \frac{{{D_S}}}{\Delta }} \right)y} \right|\!\!\begin{array}{*{20}{c}}
    {\left( {{N_T},1} \right)} \\ 
    {\left( {{N_T} - m,1} \right)} 
    \end{array}} \!\!\right)
    \notag\\
    & = \frac{{{\Delta ^{m - 1}}}}{{2\pi i}{\left( {y{D_S}} \right)}^{m-{N_T} }}
\int_\mathcal{L} {\frac{{\Gamma\!\left( { - t} \right)}}{{\Gamma\!\left( {m - t} \right)}}} {\left( {\left( {{T_S} + T_P^{\left( S \right)} + \frac{{{D_S}}}{\Delta }} \right)y} \right)^t}{\rm{d}}t,
    \end{align}
    where $ i=\sqrt{-1}$, $\mathcal{L}$ is the right loop integral path starting at the point $  - 0.1i + \infty  $, terminating at the point $  0.1i + \infty  $, and intersecting the $x$-axis at $\left( {{x_s}{\text{,0}}} \right)$, where $0>{x_s}>-1$.
    Substituting \eqref{1243ehk} into $I$, we have
    \begin{align}
    &I =  \sum\limits_{m = 1}^{{N_A}} {u\left( r_m \right)} \!\!\left(\!\!\!\! {\begin{array}{*{20}{c}}
    {{N_T} - 1} \\ 
    {m - 1} 
    \end{array}}\!\!\!\! \right)\!{\left(\! {\frac{{{D_S}}}{\Delta }} \!\right)^{{N_T} - m}}\frac{1}{{2\pi i}}\notag\\
&\times
    \int_\mathcal{L} {\frac{\Gamma\!\left( { - t} \right)}{{\Gamma\!\left( {m - t} \right)}}} {\left( {\left( {{T_S} + T_P^{\left( S \right)} + \frac{{{D_S}}}{\Delta }} \right)} \right)^t}{I_2}{\rm{d}}t,
    \end{align}
    where ${I_2}{\text{ = }}\int_0^{{a_n}} {\frac{{{y^{{N_T} - m + t}}}}{{{{\left( {1 - \frac{{\Delta {T_S} + \Delta T_P^{\left( S \right)}}}{\Delta }y} \right)}^{{N_T} - 1}}}}{\text{d}}y}$. With the help of \cite[eq. (3.194.1)]{gradshteyn2007} and \cite[eq. (9.113)]{gradshteyn2007}, $I_2$ can be solved. Substituting $I_2$ into $I$, we obtain \eqref{skjf} to finish the proof. {\color{black}The Fox’s $H$-function in \eqref{skjf} can be calculated efficiently using a Python implementation~\cite{alhennawi2015closed}.}
    \section{Proof of Theorem~\ref{them2}}\label{them2a}
    \renewcommand{\theequation}{D-\arabic{equation}}
    \setcounter{equation}{0}
    The Lagrangian function associated with problem \eqref{pro1} is given by
    \begin{align}
    F_L =& - \sum\limits_{n = 1}^{{N_T}} {\sum\limits_{m = 1}^{{N_A}} {{r_m}\mathcal{F}\left( {{a_n},{N_T},m,{D_S},{T_S},T_P^{\left( S \right)},\Delta } \right)} }
    \notag\\
    &- \lambda \left( {\sum\limits_{m = 1}^{{N_A}} {{r_m}}  - r} \right),
    \end{align}
    where $\lambda$ is the Lagrange multipliers. The Karush–Kuhn–Tucker (KKT) optimality conditions for the optimal solution are
    \begin{equation}\label{con1}
    \left\{ \begin{gathered}
    \sum\limits_{m = 1}^{{N_A}} {{r_m}}  = r \hfill \\
    \sum\limits_{n = 1}^{{N_T}} {\mathcal{F}\left( {{a_n},{N_T},m,{D_S},{T_S},T_P^{\left( S \right)},\Delta } \right)}  = \lambda ,\forall m. \hfill \\ 
    \end{gathered}  \right.
    \end{equation}
    Note that if $ {m_1} \ne {m_2} $, we have
    \begin{align}
    &\sum\limits_{n = 1}^{{N_T}} {\mathcal{F}\left( {{a_n},{N_T},{m_1},{D_S},{T_S},T_P^{\left( S \right)},\Delta } \right)}
   \notag\\
   &\ne \sum\limits_{n = 1}^{{N_T}} {\mathcal{F}\left( {{a_n},{N_T},{m_2},{D_S},{T_S},T_P^{\left( S \right)},\Delta } \right)}.
    \end{align}
    Thus, the second condition in \eqref{con1} is only correct when $N_A = 1$. Then, using the first condition, we obtain that $r_1 = r$, which completes the proof.
    
    \section{Proof of Theorem~\ref{them3}}\label{them3a}
    \renewcommand{\theequation}{E-\arabic{equation}}
    \setcounter{equation}{0}
    When the transmitters are risk averse, the Lagrangian function can be expressed as
    \begin{align}
    {F_L} = & - \sum\limits_{n = 1}^{{N_T}} {\sum\limits_{m = 1}^{{N_A}} {\ln {r_m}{\cal F}\left( {{a_n},{N_T},m,{D_S},{T_S},T_P^{\left( S \right)},\Delta } \right)} } 
\notag\\
&- \lambda \left( {\sum\limits_{m = 1}^{{N_A}} {{r_m}}  - r} \right).
    \end{align}
    The KKT conditions then can be written as
    \begin{equation}\label{aejf}
    \left\{\!\!\! \begin{array}{l}
    \sum\limits_{n = 1}^{{N_T}}\!\!{\frac{1}{{{r_m}}}{\cal F}\!\left(\! {{a_n},{N_T},m,{D_S},{T_S},T_P^{\left( S \right)},\Delta } \!\right)}  = \lambda ,\forall m\\
    \sum\limits_{m = 1}^{{N_A}} {{r_m}}  = r.
    \end{array} \right.
    \end{equation}
    By solving \eqref{aejf}, we have
    \begin{equation}\label{fale}
    \frac{1}{\lambda }\sum\limits_{n = 1}^{{N_T}} {{\cal F}\left( {{a_n},{N_T},m,{D_S},{T_S},T_P^{\left( S \right)},\Delta } \right)}  = {r_m},
    \end{equation}
    and
    \begin{equation}\label{effae}
    \lambda  = \frac{{\rm{1}}}{r}\sum\limits_{m = 1}^{{N_A}} {\sum\limits_{n = 1}^{{N_T}} {{\cal F}\left( {{a_n},{N_T},m,{D_S},{T_S},T_P^{\left( S \right)},\Delta } \right)} }.
    \end{equation}
    Substituting \eqref{effae} into \eqref{fale}, we can derive \eqref{afeaf} to complete the proof.
    \end{appendices}
    \bibliographystyle{IEEEtran}
    \bibliography{IEEEabrv,Ref}
    \end{document}